\documentclass{emulateapj}
\usepackage{appendix}
\usepackage{color}
\newcommand{\cgsflux}{erg~s$^{-1}$~cm$^{-2}$}

\newcommand{\ergss}{erg~s$^{-1}$}
\newcommand{\kms}{\hbox{km~s$^{-1}$}}

\newcommand{\ArII}{\hbox{{\rm Ar}\kern 0.1em{\sc ii}}}
\newcommand{\ArIII}{\hbox{{\rm Ar}\kern 0.1em{\sc iii}}}
\newcommand{\CIV}{\hbox{{\rm C}\kern 0.1em{\sc iv}}}
\newcommand{\HI}{\hbox{{\rm H}\kern 0.1em{\sc i}}}
\newcommand{\HII}{\hbox{{\rm H}\kern 0.1em{\sc ii}}}
\newcommand{\HeI}{\hbox{{\rm He}\kern 0.1em{\sc i}}}
\newcommand{\HeII}{\hbox{{\rm He}\kern 0.1em{\sc ii}}}
\newcommand{\NII}{\hbox{{\rm N}\kern 0.1em{\sc ii}}}
\newcommand{\OI}{\hbox{{\rm O}\kern 0.1em{\sc i}}}
\newcommand{\OII}{\hbox{{\rm O}\kern 0.1em{\sc ii}}}
\newcommand{\OIII}{\hbox{{\rm O}\kern 0.1em{\sc iii}}}
\newcommand{\OIIlong}{{\rm O}\kern 0.1em{\sc ii}~$\lambda 3727$} 
\newcommand{\FeII}{\hbox{{\rm Fe}\kern 0.1em{\sc ii}}}
\newcommand{\NeII}{\hbox{{\rm Ne}\kern 0.1em{\sc ii}}}
\newcommand{\NeIII}{\hbox{{\rm Ne}\kern 0.1em{\sc iii}}}
\newcommand{\NeV}{\hbox{{\rm Ne}\kern 0.1em{\sc v}}}
\newcommand{\SII}{\hbox{{\rm S}\kern 0.1em{\sc ii}}}
\newcommand{\SIII}{\hbox{{\rm S}\kern 0.1em{\sc iii}}}
\newcommand{\SIV}{\hbox{{\rm S}\kern 0.1em{\sc iv}}}
\newcommand{\SiIV}{\hbox{{\rm Si}\kern 0.1em{\sc iv}}}
\newcommand{\MgII}{\hbox{{\rm Mg}\kern 0.1em{\sc ii}}}
\newcommand{\Halpha}{\hbox{{\rm H}\kern 0.1em$\alpha$}}
\newcommand{\Hbeta}{\hbox{{\rm H}\kern 0.1em$\beta$}}
\newcommand{\Heopta}{\hbox{{\rm He}\kern 0.1em{\sc i}}~$6678$}
\newcommand{\Heoptb}{\hbox{{\rm He}\kern 0.1em{\sc i}}~$5876$}
\newcommand{\Heoptc}{\hbox{{\rm He}\kern 0.1em{\sc i}}~$4471$}
\newcommand{\Brgam}{\hbox{{\rm Br}\kern 0.1em$\gamma$}}
\newcommand{\Brten}{\hbox{{\rm Br}\kern 0.1em$10$}}
\newcommand{\Breleven}{\hbox{{\rm Br}\kern 0.1em$11$}}
\newcommand{\HeIh}{\hbox{{\rm He}\kern 0.1em{\sc i}}~$1.7$~{\micron}}
\newcommand{\HeIk}{\hbox{{\rm He}\kern 0.1em{\sc i}}~$2.06$~{\micron}}
\newcommand{\squishlist}{
   \begin{list}{$\bullet$}
    { \setlength{\itemsep}{0pt}      \setlength{\parsep}{1pt}
      \setlength{\topsep}{3pt}       \setlength{\partopsep}{0pt}
      \setlength{\leftmargin}{1.5em} \setlength{\labelwidth}{1em}
      \setlength{\labelsep}{0.5em} } }
\newcommand{\squishend}{
    \end{list}  }

\received{}
\accepted{}
\slugcomment{Accepted for publication in ApJ}
\shorttitle{{\it NuSTAR} Survey of Local ULIRGs}
\shortauthors{Teng et al.}

\begin{document}

\title{A {\it NuSTAR} Survey of Nearby Ultraluminous Infrared Galaxies}
\author{Stacy H. Teng\altaffilmark{1, 2, 3}, Jane R. Rigby\altaffilmark{1}, 
Daniel Stern\altaffilmark{4}, 
Andrew Ptak\altaffilmark{1},
D.~M.~Alexander\altaffilmark{5},  
Franz~E.~Bauer\altaffilmark{6,7,8}, 
Stephen E. Boggs\altaffilmark{9}, 
 W. Niel Brandt\altaffilmark{10, 11},  
 Finn E.~Christensen\altaffilmark{12}, 
 Andrea Comastri\altaffilmark{13}, 
William~W.~Craig\altaffilmark{9,14}, 
Duncan~Farrah\altaffilmark{15}, 
Poshak~Gandhi\altaffilmark{16}, 
Charles~J.~Hailey\altaffilmark{17}, 
Fiona~A.~Harrison\altaffilmark{18}, 
Ryan~C.~Hickox\altaffilmark{19}, 
Michael~Koss\altaffilmark{20}, 
Bin~Luo\altaffilmark{10,11}, 
Ezequiel~Treister\altaffilmark{21}, 
and William~W.~Zhang\altaffilmark{1}
}

\altaffiltext{1}{Astrophysics Science Division, NASA Goddard  Space Flight Center, Greenbelt, MD 20771, USA}
\altaffiltext{2}{Department of Astronomy, University of Maryland, College Park, MD 20742, USA}
\altaffiltext{3}{Current address: Science and Technology Division, Institute for Defense Analyses, 4850 Mark Center Drive, Alexandria, VA 22311, USA}
\altaffiltext{4}{Jet Propulsion Laboratory, California Institute of Technology, Pasadena, CA 91109, USA}
\altaffiltext{5}{Department of Physics, Durham University, Durham, DH1 3LE, UK}
\altaffiltext{6}{Instituto de Astrof\'{\i}sica, Facultad de F\'{i}sica, Pontificia Universidad Cat\'{o}lica de Chile, 306, Santiago 22, Chile} 
\altaffiltext{7}{Millennium Institute of Astrophysics, Santiago, Chile} 
\altaffiltext{8}{Space Science Institute, 4750 Walnut Street, Suite 205, Boulder, Colorado 80301}
\altaffiltext{9}{Space Sciences Laboratory, University of California, Berkeley, CA 94720, USA}
\altaffiltext{10}{Department of Astronomy and Astrophysics, The Pennsylvania State University, 525 Davey Lab, University Park, PA 16802, USA}
\altaffiltext{11}{Institute for Gravitation and the Cosmos, The Pennsylvania State University, University Park, PA 16802, USA}
\altaffiltext{12}{DTU Space-National Space Institute, Technical University of Denmark, Elektrovej 327, DK-2800 Lyngby, Denmark}
\altaffiltext{13}{INAF-Osservatorio Astronomico di Bologna, via Ranzani 1, I-40127 Bologna, Italy}
\altaffiltext{14}{Lawrence Livermore National Laboratory, Livermore, CA 94550, USA}
\altaffiltext{15}{Department of Physics, Virginia Tech, Blacksburg, VA 24061, USA}
\altaffiltext{16}{School of Physics \& Astronomy, University of Southampton, Highfield, Southampton, SO17 1BJ, UK}
\altaffiltext{17}{Columbia Astrophysics Laboratory, Columbia University, New York, NY 10027, USA}
\altaffiltext{18}{Cahill Center for Astronomy and Astrophysics, California Institute of Technology, Pasadena, CA 91125, USA}
\altaffiltext{19}{Department of Physics and Astronomy, Dartmouth College, 6127 Wilder Laboratory, Hanover, NH 03755, USA}
\altaffiltext{20}{Institute for Astronomy, Department of Physics, ETH Zurich, Wolfgang-Pauli-Strasse 27, CH-8093 Zurich, Switzerland}
\altaffiltext{21}{Departamento de Astronom\'{\i}a Universidad de Concepci\'{o}n, Casilla 160-C, Concepci\'{o}n, Chile} 

\begin{abstract}
We present a {\it NuSTAR}, {\it Chandra}, and {\it XMM--Newton} survey of nine of the 
nearest ultraluminous infrared galaxies (ULIRGs).  The unprecedented sensitivity of
{\it NuSTAR} at energies above 10~keV enables spectral modeling with far better precision
than was previously possible.  Six of the nine sources observed were
detected sufficiently well by {\it NuSTAR} to model in detail their broadband X-ray spectra,
and recover the levels of obscuration and intrinsic X-ray luminosities.  Only one source
(IRAS~13120--5453) has a spectrum consistent with a Compton--thick AGN, but we cannot rule out that a second source (Arp 220) harbors an extremely highly obscured 
AGN as well.  Variability in column density (reduction by a factor of a few compared to older observations) is seen in IRAS~05189--2524 and Mrk~273, altering the classification of these border-line sources from Compton-thick to Compton-thin.  
The ULIRGs in our sample have surprisingly low observed fluxes in high energy ($>$10~keV) X-rays, 
especially compared to their bolometric luminosities.  
They have lower ratios of unabsorbed 2--10~keV to bolometric luminosity, and
unabsorbed 2--10~keV to mid-IR [O IV] line luminosity than do Seyfert 1 galaxies.
We identify IRAS~08572+3915 as another candidate
intrinsically X-ray weak source, similar to Mrk~231.  
We speculate that the X-ray weakness of IRAS~08572+3915 is related 
to its powerful outflow observed at other wavelengths.  
\end{abstract}


\section{Introduction}
\label{sec:intro}

 In 1984, the {\it Infrared Astronomical Satellite} ({\it IRAS}) identified a
 large population of ``ultra-luminous infrared galaxies'' (ULIRGs), 
which are as luminous as quasars\footnote{ULIRGs are defined as galaxies
whose 8--1000~\micron\ luminosity is $10^{12}$ to $10^{13}$~$L_\odot$.},
but whose power emerges almost entirely in the infrared \citep{neugebauer84,aaronson84}.
The {\it Infrared Space Observatory} ({\it ISO}) and the {\it Spitzer Space Telescope} 
have subsequently shown that galaxies with ULIRG luminosities 
are critical for building  the stellar mass of galaxies:  they are a
thousand times more common at significant redshift than today, 
and the luminosity function evolves so steeply that  at $z>2$ the 
bulk of star formation occurs
in galaxies with ULIRG luminosities\citep{lefloch05,pg05,caputi07}.  
It is worth noting that galaxies with ULIRG luminosities at $z\ga1$
where such high luminosities are common, 
may be very different than ULIRGs at $z=0$, where such high luminosities are rare 
\citep{Rowan-Robinson04, Rowan-Robinson05, Sajina06, Papovich07, Rigby08, Farrah08, 
MenendezDelmestre09, Symeonidis09, Elbaz10, Hwang10, Rujopakarn11, Gladders13}.
Bearing that caveat in mind, it is still highly instructive to study ULIRGs at $z=0$,
since these nearby galaxies can be studied far more comprehensively than can ULIRGs 
in the distant universe.

Since ULIRGs in the nearby universe are generally merging galaxies with active
galactic nuclei (AGN) signatures, they have inspired
an evolutionary paradigm in which massive, gas-rich galaxies collide, rapidly form stars, 
feed a buried AGN, and then shine as an unobscured quasar \citep{sanders88, genzel, kim02}.  
In simulations \citep{springel, dimatteo, hopkins08}, during the final coalescence of 
the merging galaxies, massive gas inflows trigger rates of star formation as high as 
those inferred for ULIRGs, and  nuclear accretion may be obscured
by large, even Compton-thick ($N_H\gtrsim10^{24}~{\rm cm}^{-2}$)
column densities. 
In this picture, feedback from nuclear accretion eventually disperses the gas,
and a traditional optical quasar is revealed.
Supporting this picture is the observational evidence that AGN are more common in 
ULIRGs with morphologies that are advanced mergers \citep[e.g.,][]{vei09a, teng10}. 

It is one thing to find AGN signatures in a galaxy, and a very different thing
to find that the AGN dominates the galaxy's energy production.
For a quarter century, a key question has been, 
``Are ULIRGs powered mainly by star formation, or by nuclear accretion?''
\citep[e.g.,][]{genzel98, armus, farrah07, vei09a}.  
Excitation diagrams \citep{genzelcesarsky00, armus} using mid-infrared diagnostics
were developed to determine which ULIRGs are AGN dominated.  
While these diagnostics generally correlate, systematic errors and considerable scatter 
remain \citep{vei09a}.
Thus, while excitation diagrams have indicated which ULIRGs have larger AGN contributions
in a relative sense, they have not settled the question of whether accretion
power dominates the energetics of ULIRGs.

Many X-ray surveys have attempted to quantify the AGN contribution in ULIRGs.
For example, half of the ULIRG sample of \citet{koss13} was detected by the {\it Swift} Burst Alert Telescope (BAT) 
at energies above 14~keV.
A complication is that ULIRGs are notoriously X-ray weak, generally believed to be due to
obscuration \citep[e.g.,][]{ptak03, frances03, teng05}.

There is growing evidence that the AGN in many ULIRGs are obscured by Compton-thick column 
densities.  For example, IRAS~F04103-2838 has an iron line with a large equivalent
  width (EW) of $\sim 1.6$~keV \citep{teng08}, as predicted for AGN obscured
by high column densities \citep{KrolikKallman87,Levenson02}.  
As a second example, 
using {\it Suzaku}, 
\citet{braito09} reported a direct AGN component above 10~keV in the Superantennae, 
and  \citet{teng09} reported a marginal detection in Mrk~273.
However, these latter detections are at very low levels, near the sensitivity limits of 
{\it Suzaku} and {\it BeppoSAX}, and may additionally suffer from contamination 
by unrelated sources due to the limited spatial 
resolution of those observatories at energies above $10$~keV.  

An example of these limitations can be found in the case of Mrk~231:
{\it BeppoSAX} and {\it Suzaku} reported a direct AGN component above 10~keV 
\citep[]{braito, piconcelli}.  
However, recent {\it Nuclear Spectroscopic Telescope Array} ({\it NuSTAR}; \citealt{nustar})
observations of Mrk~231 found that the AGN is in fact
intrinsically X-ray weak rather then highly obscured as previously
thought \citep{teng14}.  The ratio of intrinsic 2-10 keV luminosity to bolometric luminosity 
for Mrk~231 is
only 0.03\%, compared to the 2--15\% seen in Seyferts and radio-quiet
quasars \citep{elvis}.  For objects accreting at close to the Eddington rate, the same ratio is typically 
$\sim$0.3--0.7\% \citep[e.g.,][]{lusso10, lusso12, vf09}.   
Thus, the previous claimed detections at energies above 10~keV may have been due to contamination \citep{teng14}.  

{\it NuSTAR} brings improved angular resolution
(half-power diameter, or HPD, $\sim$58$\arcsec$) and improved 
sensitivity at energies above 10~keV to bear on the problem of whether 
AGN contribute significantly to the bolometric output of ULIRGs.
In this paper, we present the results of a {\it NuSTAR} survey of nine
of the nearest ULIRGs.
This paper is organized as follows:  \S2 details our sample and their
multi-wavelength properties; \S3 presents the new X-ray observations
obtained for this study; \S4 presents the {\it NuSTAR} photometry
of our sample; \S5 presents a detailed broadband X-ray spectral analysis of our
sample, \S6 discusses the general properties of ULIRGs in our study; and \S7 summarizes our results.  Throughout this paper, we adopt $H_0$ = 71 km s$^{-1}$ Mpc$^{-1}$, $\Omega_M = 0.27$, and $\Lambda = 0.73$ \citep{cosmo}.  
Luminosities taken from the literature have been recalculated for our assumed cosmology.  

\section{The Sample}
\label{sec:sample}

\subsection{The Selection}

During its two-year baseline mission, {\it NuSTAR} observed a sample
of nine of the nearest ($z < 0.078$) ULIRGs, out of the total
sample of $\sim 25$ ULIRGs within that volume based on the selection
of \citet{rbgs}.  The sample was split into a ``deep'' sample of five
ULIRGs observed for  $>50$~ks, and a ``shallow''
sample of four objects observed for $\sim 25$~ks.  The deep sample
consists of the four ULIRGs at $z < 0.078$ which are brightest at
$60 \mu$m.  This selection by low redshift and high infrared
brightness was done to maximize the likelihood of obtaining
high-quality {\it NuSTAR} data. 
The Superantennae was added to the deep sample because 
previous observations at $>10$~keV suggested the presence 
of a Compton-thick AGN.
All five deep survey sources have previous observations above 10~keV
by {\it BeppoSAX} or {\it Suzaku} PIN; three have reported detections.
In addition, the deep sample targets have simultaneous
soft X-ray coverage from either {\it Chandra}
(Mrk~231; \citealt{teng14}) or {\it XMM-Newton} (the other four).  
 Table~\ref{tab:sample} lists the targets and exposure times for the entire sample.
The shallow sample targeted nearby ULIRGs showing  AGN signatures in their optical spectra.  
Thus, the shallow sample,  like the deep sample, is biased toward high detection rates at X-ray energies.

\subsection{Multiwavelength Characterization of the NuSTAR Sample}

The targets in our sample have been very well studied across the
electromagnetic spectrum.  In this section, we note the information at
other wavelengths and past X-ray observations that are relevant to our study.  

\subsubsection{The Deep Survey}

\begin{itemize}

\item
IRAS~05189--2524 is an advanced--stage merger with a single nucleus.
Optically classified as a Seyfert 2, near-infrared
spectroscopy reveals the presence of a hidden broad line region
via broad Paschen $\alpha$ \citep{vei99a, vei99b}. 
Continuum and emission-line diagnostics from {\it Spitzer} spectra indicate 
that the infrared luminosity of this source is dominated by an AGN \citep{vei09a}.

In the X-ray band, IRAS~05189--2524 is one of the brightest ULIRGs on the sky.
Historic {\it XMM-Newton} and {\it Chandra} data imply a 2--10~keV X-ray continuum with
a luminosity of $\sim$10$^{43}$~erg~s$^{-1}$.   The source was observed
by {\it Suzaku} in 2006, at which time its 2--10~keV flux appears to have
dropped by a factor of $\sim$30 and the Fe line became more prominent.
Its 0.5--2~keV flux appeared unchanged.  The target was undetected by
{\it Suzaku} PIN \citep{teng09}, and those data were unable to distinguish
whether this change in observed flux was due to a change in the
intrinsic AGN luminosity or to a change in the thickness of the absorbing column.
IRAS~05189--252 was also detected by {\it Swift} BAT 
in the 14--195~keV energy range \citep{koss13}. Of the four ULIRGs 
surveyed by \citet{koss13}, IRAS~05189--2524 has the most significant detection in the 
24--35~keV band, at 4.2~$\sigma$.

\item 
Mrk 231 is a merger remnant that contains both an intense starburst and a 
luminous quasar with a Type 1 optical spectral classification.  
It is also a rare iron low-ionization broad absorption line quasar \citep[FeLoBAL; e.g.,][]{aw72, g02, g05, vei13a}.
Continuum and emission-line diagnostics from {\it Spitzer} spectra indicate 
that the infrared luminosity of this source is dominated by an AGN \citep{vei09a}.
\citet{teng14} found that the AGN is Compton-thin and intrinsically 
X-ray weak, with the intrinsic 2--10~keV luminosity  being only 0.03\%
of the AGN bolometric luminosity. 
The \textit{NuSTAR} and \textit{Chandra}  X-ray data on this source were also 
analyzed by \citet{feruglio15}.  Their results suggest the presence of an ultra-fast outflow
where the ionized wind reaches speeds of $\sim2\times 10^{4}$~\kms.

\item 
Mrk 273 shows a single nucleus in UV and optical images, but the near-infrared
reveals a double nucleus \citep{armus90, surace00, scoville00}.  
It is optically classified as a Seyfert~2 galaxy
\citep{khachikian74, vei99a}.  Continuum and emission-line
diagnostics from {\it Spitzer} spectra indicate 
that roughly half the infrared luminosity of this source is powered by an AGN \citep{vei09a}.

\citet{teng09} reported a  1.8$\sigma$ detection of Mrk 273 by {\it Suzaku} PIN.
Their best-fit model used two partial covering absorbers to
model minor spectral variability below 10 keV over six years.  The
model favored the scenario in which the covering fractions of the
absorbers ($N_{H,1} \sim 1.6 \times 10^{24}$~cm$^{-2}$, $N_{H,2} \sim 3\times
10^{23}$~cm$^{-2}$) was time-variable.  
\citet{koss13} reported a 2.4$\sigma$ detection in the 24--35~keV band.

\item
Arp 220 is the closest ULIRG, has a double nucleus, and is 
one of the most famous infrared sources.  It is optically 
classified as a LINER \citep{armus89, taniguchi99}.
CO observations suggest that the western nucleus hosts a deeply buried AGN.
The total column density of that nucleus is $\sim 10^{25}$ cm$^{-2}$ \citep{downes07,scoville14}.
Continuum and emission-line diagnostics from {\it Spitzer} spectra
indicate that the AGN powers a small fraction of the total infrared
luminosity \citep{armus, vei09a}.

Past X-ray observations find that the nuclear spectrum of Arp 220 is best-fit by
a flat power law \citep{ptak03} and \citet{iwasawa05} detected a
strong Fe K emission line (EW$=1.9 \pm 0.9$~keV) in  low--quality {\it XMM-Newton}
data, implying a Compton-thick nucleus.  The detection of the
  line was confirmed by \citet{teng09} using {\it Suzaku} data, but with an EW of
  only $0.42^{+0.54}_{-0.32}$ ~keV.  Arp 220  was undetected by  {\it Suzaku} 
above 10~keV \citep{teng09}.

\item
The Superantennae, also known as IRAS~F19254-7245, 
is a binary ULIRG whose southern nucleus is optically classified as
  a Type 2 AGN \citep{degrijp}.
Its {\it XMM-Newton} spectrum shows a
  hard power-law continuum above 2~keV ($\Gamma = 1.3$) and has an Fe
  line with an EW of 1.4 keV \citep{braito03}.  It was   marginally detected above 10~keV 
  by {\it Suzaku} \citep{braito09}.   The {\it Suzaku} data indicate that the
  Superantennae harbors a Compton-thick AGN with a column density of
  $\sim 3\times 10^{24}$~cm$^{-2}$.  \citet{jia12} found that the
  relative line strength between the 6.4 and 6.7~keV lines 
as seen by \textit{Chandra} varied  between 2001 and 2009.  

\end{itemize}

\subsubsection{The Shallow Survey}

\begin{itemize}

\item IRAS~08572+3915 is a double nucleus ULIRG whose northwestern nucleus is
  thought to host an AGN.  Its optical spectrum is intermediate
  between a LINER and Seyfert~2 \citep{vei95, vei99a}.
 Continuum and emission-line diagnostics
  from {\it Spitzer} spectra indicate that the infrared luminosity of this
  source is dominated by an AGN \citep{vei09a}.

No significant X-ray detection of this
  source has been reported, but {\it Chandra} data show a detection of a
  few counts.  Using the hardness ratio between the 0.5--2 and 2--8~keV {\it Chandra} bands 
and assuming a power-law spectrum with $\Gamma=1.8$, 
\citet{teng09} estimated a 0.5--10~keV flux of $\sim
  3\times10^{-14}$ erg~s$^{-1}$~cm$^{-2}$, corresponding to a
  luminosity of $\sim 2 \times 10^{41}$~erg~s$^{-1}$.  The target was previously observed, but was not detected by  {\it Suzaku} \citep{teng09}.

\item IRAS~10565+2448 is a pair of interacting spiral galaxies, with two distinct
nuclei in the optical and near-infrared \citep{scoville00}.  
The western nucleus is much brighter.
The optical spectrum is
that of an H~II region or a LINER \citep{vei95, vei99a}.
Continuum and emission-line diagnostics from {\it Spitzer} spectra indicate 
that the infrared luminosity of this source is dominated by star formation, not an 
AGN \citep{vei09a}.  This source was detected by both {\it Chandra}
and {\it XMM-Newton}.  These spectra were typical of ULIRGs, 
with a power law component as well  as a MEKAL hot gas component \citep[e.g.,][]{teng10}.
In two observations, the 2--10~keV flux was 
$3.7^{+2}_{-4}  \times 10^{-14}$ \cgsflux\ and 
$6.7^{+1.7}_{-2.8} \times 10^{-14}$ \cgsflux, 
the inferred intrinsic 2--10~keV luminosity 
was $1.7\times10^{41}$~erg~s$^{-1}$ and 
    $3.7\times10^{41}$~erg~s$^{-1}$,
the spectral index of the best-fit model was 
$\Gamma = 2.14^{+0.66}_{-0.53}$ and $1.41^{+0.25}_{-0.23}$, 
and the $kT$ MEKAL temperature was 
$0.68^{+0.17}_{-0.12}$~keV and $0.68^{+0.14}_{-0.07}$~keV
\citep{teng10}.

\item IRAS~13120--5453 
has a morphology that is classified as ``single or obscured nucleus with 
long prominent tails'' by \citet{haan11}.  Its optical spectral classification
is Seyfert 2 \citep{vv2010}.
This source was detected by {\it Chandra} \citep{iwasawa11} and has extended soft X-ray emission.
The observed 2--7 keV band flux is $1.4\times 10^{-13}$\cgsflux\ and the 
2--10~keV X-ray luminosity assuming no extinction other than Galactic is 
$4.5 \times 10^{41}$~\ergss.
The 3--7 keV spectrum is fit by a power law of $\Gamma = 2.6^{+1.5}_{-0.9}$,
and the 0.4--2~keV data are fit with a MEKAL component of $kT = 0.82^{+0.26}_{0.14}$~keV 
\citep{iwasawa11}.

\item IRAS~14378--3651   %
shows a single nucleus \citep{bushouse2002} in optical and near-infrared {\it HST} images.
It is optically classified as a LINER \citep{kim98}.
This source was detected by {\it Chandra} with 
40 counts in 14~ks \citep{iwasawa11}. 
The observed 2--7 keV band flux is $2.1\times 10^{-14}$\cgsflux\ and the 
2--10~keV X-ray luminosity assuming no extinction other than Galactic is 
$3.4 \times 10^{41}$~\ergss.
Its hardness ratio implies a  highly absorbed spectrum with $\Gamma \sim 0.35$.

\end{itemize}

\begin{turnpage}
\begin{deluxetable*}{cccccccccccccc}
\tablecolumns{13}
\tabletypesize{\scriptsize}
\setlength{\tabcolsep}{0.01in}
\tablecaption{The Sample}
\tablewidth{0pt}
\tablehead{\colhead{Source} & \colhead{$z$} & \colhead{log $L_{\rm bol}$} & \colhead{f$_{\nu}$(60 $\mu$m)} &
  \colhead{Spectral} &\colhead{Interaction}&\colhead{\%
    AGN}&\colhead{$N_{\rm H, Gal}$}&\colhead{Previous Obs}&\colhead{\it{NuSTAR}}&\colhead{\it{NuSTAR}}&\colhead{\it{NuSTAR}}&\colhead{XMM}&\colhead{XMM}\\
\colhead{Name} & \colhead{}&\colhead{[L$_\odot$]}& [Jy]&\colhead{Type}
&\colhead{Class}&\colhead{}&\colhead{[10$^{20}$~${\rm cm}^{-2}$]}&\colhead{$>$
  10~keV}&\colhead{Obs Date}&\colhead{ObsID}&\colhead{GTI [ks]}&\colhead{ObsID}&\colhead{GTI [ks]}\\
\colhead{(1)}&\colhead{(2)}&\colhead{(3)}&\colhead{(4)}&\colhead{(5)}&\colhead{(6)}&\colhead{(7)}&\colhead{(8)}&\colhead{(9)}&\colhead{(10)}&\colhead{(11)}&\colhead{(12)}&\colhead{(13)} & \colhead{14}
}
\startdata
\cutinhead{Deep Survey}\\
IRAS 05189--2524&0.043&12.22&14&S2&IVb&71.3&1.92&PIN&2013 Feb 20&60002027002&21.3&\nodata&\nodata\\
&&&&&&&&&2013 Oct 2&60002027003&25.4&0722610101&30.8\\
&&&&&&&&&2013 Oct 2&60002027004&8.2&\nodata&\nodata\\
Mrk 231&0.042&12.60&31&S1&IVb&70.9&1.26&SAX, PIN&2012 Aug 27&60002025002&41.1&\nodata&\nodata\\
&&&&&&&&&2013 May 9&60002025004&28.6&\nodata&\nodata\\
Mrk 273&0.038&12.24&22.5&S2&IVb&45.8&1.09&PIN&2013 Nov 4&60002028002&69.9&0722610201&4.2\\
Arp 220&0.018&12.26&104&L&IIIb&18.5&4.27&PIN&2013 Aug 13 &60002026002&66.8&0722610301&29.5\\
Superantennae&0.062&12.10&5.5&S2&\nodata&41.8\tablenotemark{a}&5.90&SAX, PIN&2013 May
26&60002029002&58.7&\nodata&\nodata\\
&&&&&&&&&2013 Sept 2&60002029004&31.0&0722610401&29.4\\
\cutinhead{Shallow Survey}\\
IRAS 08572+3915&0.058&12.22&7.4&L/S2&IIIb&71.6&2.60&PIN&2013 May 23&60001088002&24.1&\nodata&\nodata\\
IRAS 10565+2448&0.043&12.11&12&HII/L&\nodata&16.6&1.54&\nodata&2013 May 22&60001090002&25.4&\nodata&\nodata\\
IRAS 13120--5453&0.031&12.45&41&S2&\nodata&17.3\tablenotemark{a}&26.10&\nodata&2013 Feb
25&60001091002&26.2&\nodata&\nodata\\
IRAS 14378--3651&0.068&12.33&6&L/S2&\nodata&14.6\tablenotemark{a}&1.22&\nodata&2013
Feb 28&60001092002&24.5&\nodata&\nodata\\
\enddata
\tablenotetext{a}{These three objects were not studied in
  \citet{vei09a}.  To calculate the AGN fraction in these sources, we
  downloaded reduced spectra from the Cornell Atlas of
  {\it Spitzer}/Infrared Spectrograph Sources \citep[CASSIS;][]{cassis}.
  We then fit each spectrum using PAHFIT version 1.2 \citep{pahfit}.
  These measurements were used to estimate the AGN fraction following
  the recipe for methods 1--4 and 6 in \citet{vei09a}.  The estimated
  AGN fraction from these five methods were averaged together.}
\tablecomments{
Col.(1): Source name.  Col.(2):  Redshift.  Col.(3): Bolometric
luminosity which is assumed to be 1.15L$_{8-1000 \mu m}$.  Col.(4): Optical spectral type 
(S1 $=$ Seyfert 1; S2 $=$ Seyfert 2; L $=$ LINER; HII $=$ H~II region).  
Col.(5):  Interaction class, which is a proxy for the age of the merger, ranging from III to V
(See Veilleux et al. 2002).  
Col.(6) \textit{IRAS} 60~$\mu$m flux density from \citet{ks98}.
Col.(7):  The fraction of the total infrared
luminosity attributed to the AGN using six independent methods of
evaluating line and continuum {\it Spitzer} data (Veilleux et
al. 2009).  Col.(8):  The Galactic column density from Dickey and
Lockman (1990).  Col.(9): Previous pointed hard X-ray observations.  Col.(10):
Observation start date of the {\it NuSTAR} data.  Col.(11): {\it NuSTAR} observation 
identification.  Col.(12):  {\it NuSTAR} good time
interval.  Col.(13): Simultaneous {\it XMM-Newton} observation
identification.  Col.(14): {\it XMM-Newton} good time interval.
}
\label{tab:sample}
\end{deluxetable*}
\end{turnpage}

\section{Observations and Data Reduction}
\label{sec:obs}

The targets in our sample were observed by {\it NuSTAR} between 2012
August and 2013 November.  {\it NuSTAR} observed the Deep Survey
targets for a total of 50--100 ks per target over one or more epochs.  
These objects were also observed
with {\it Chandra} \citep[Mrk 231;][]{teng14} or {\it
  XMM-Newton} (IRAS~05189--2524, Mrk~273, Arp~220, and Superantennae) to 
constrain their low energy properties  and check for spectral
variability.  The observations were designed such that one {\it NuSTAR} epoch 
for each target was simultaneous with its observation by  {\it XMM-Newton}  (PI: Teng).
 IRAS~05189--2524, Arp~220, and the Superantennae were
observed for 33~ks by {\it XMM-Newton}.  Due to visibility
constraints, the total {\it XMM-Newton} exposure on Mrk~273 was only
24~ks.   Table~\ref{tab:sample} lists the exposure times and dates for
these {\it NuSTAR} and {\it XMM-Newton} observations.

\subsection{Low Energy Coverage}
\label{sec:xmmobs}

The {\it XMM-Newton} data were obtained using the EPIC array in full
window imaging mode.  For these observations, the medium optical
blocking filter was applied.  The data were reduced using {\it
  XMM-Newton} Science analysis Software (SAS) version 13.5.0.  The
most up-to-date calibration files, as of 2014 May, were used for the
reduction.  We followed the {\it XMM-Newton} 
ABC guide\footnote{http://heasarc.gsfc.nasa.gov/docs/xmm/abc/} 
to extract images and
spectra for our Deep Survey targets.  In particular, during the data
reduction process, portions of data with high background flares were
removed.  The Mrk 273 data were highly affected by these flares; thus,
the calibrated data only contain $\sim$17\% of the original total
exposure.  All of the Deep Survey targets appear to be point sources.
In order to ensure that we are probing the same spatial scale as the
{\it NuSTAR} data, we used the same source extraction regions as {\it
  NuSTAR} data (see below).  The background spectra are extracted from
nearby source-free areas on the same chip using the same-sized
extraction regions.  The extracted spectra were binned to 15--50
counts per bin, depending on the source count rate, such that $\chi^2$
statistics may be used.

For IRAS~13120--5453, archived {\it Chandra} data were used to extract
a low-energy X-ray spectrum (PI: Sanders).  These data were reduced
using CIAO 4.5 with CALDB version 4.5.6.  The standard calibration
procedures were followed for reducing ACIS-S data in VFAINT mode using
the chandra\_repro script \footnote {http://cxc.harvard.edu/ciao/ahelp/chandra\_repro.html}.  
As with the {\it XMM-Newton} data, the same {\it NuSTAR} source region was used to
extract the source spectrum.  The background spectrum was extracted
using a same-size region in a nearby source-free area.  The spectrum
was binned to at least 15 counts per bin so that $\chi^2$ statistics
can be applied.

\subsection{{\it NuSTAR} Observations}
\label{sec:nuobs}

The {\it NuSTAR} observations were reduced using the {\it NuSTAR} Data
Analysis Software (NuSTARDAS) that is part of HEASoft version 6.15.1,
with {\it NuSTAR} calibration database version 20131223.  The script
{\it nupipeline} was used to produce calibrated event files for each
of the two focal plane modules \citep[FPMA and FPMB;][]{nustar}.
The good time intervals of these events are listed in
Table~\ref{tab:sample}.  All spectra were binned such that $\chi^2$ statistics can be used.

The $E<20$~keV {\it NuSTAR} background is spatially non-uniform over FPMA and
FPMB, caused by stray light incompletely blocked by the aperture stop.
To correct for this aperture background and the instrumental background, we followed the producedure in
\citet{wik14} to create simulated total background events for each
source.  The simulated backgrounds were scaled for exposure time, size
of the extraction region, and the response of individual chips for
each observation.   The simulated
backgrounds were used to create background-subtracted images for
photometry in Section~\ref{sec:phot} and background spectra for the spectral analysis of the 
faint sources (IRAS~13120--5453, the Superantennae, and Arp~220) in Section~\ref{sec:spec}.  We
conservatively estimate that the broadband systematic uncertainties in
the derived background spectra are $\lesssim$5\%. 

With the exception of Mrk~273, source spectra were extracted using
circular apertures with 1$\arcmin$ radii.  For Mrk~273, due to the projected
vicinity of a background source \citep[Mrk~273X;][]{mrk273x}, the
source spectrum was extracted using a circle with 0$\farcm$8
radius. Spectral analysis was performed using HEASoft version 6.15.1.
For objects with multiple observations, the spectra were co-added
using the FTOOL {\it addascaspec}.  When modeling the spectra, an additional
constant factor, typically on the order of a few percent, is applied
to account for the cross-normalization between FPMA and FPMB, and
between FPMA and {\it XMM-Newton} EPIC-pn. 
These 
cross-normalization constants were allowed to vary for 
IRAS~05189--252 and Mrk~273 since these are the brightest
sources, where more degrees of fit are possible; for the fainter 
IRAS~13120--5453, Arp~220, and the Superantennae they were held fixed.
The cross-normalization values used were current 
at the time of the modeling.  Subsequently, cross-normalization values
have been published by \citet{Madsen15} in the  \textit{NuSTAR} calibration paper.
The values we used differ by only a few percent from those of \citet{Madsen15}, 
and the difference has negligible impact on our results.

We assumed the \citet{wilms} abundances and the \citet{verner} photoelectric cross
sections in our spectral modeling with XSPEC.  The assumed column densities due
to Galactic absorption ($N_{\rm H, Gal}$) are given in Table~\ref{tab:sample}.
All errors quoted in this paper are at the 90\% confidence level
($\Delta\chi^2 = 2.706$ for a single parameter).

\begin{figure}[h]
\centering
\includegraphics[width=3.3in]{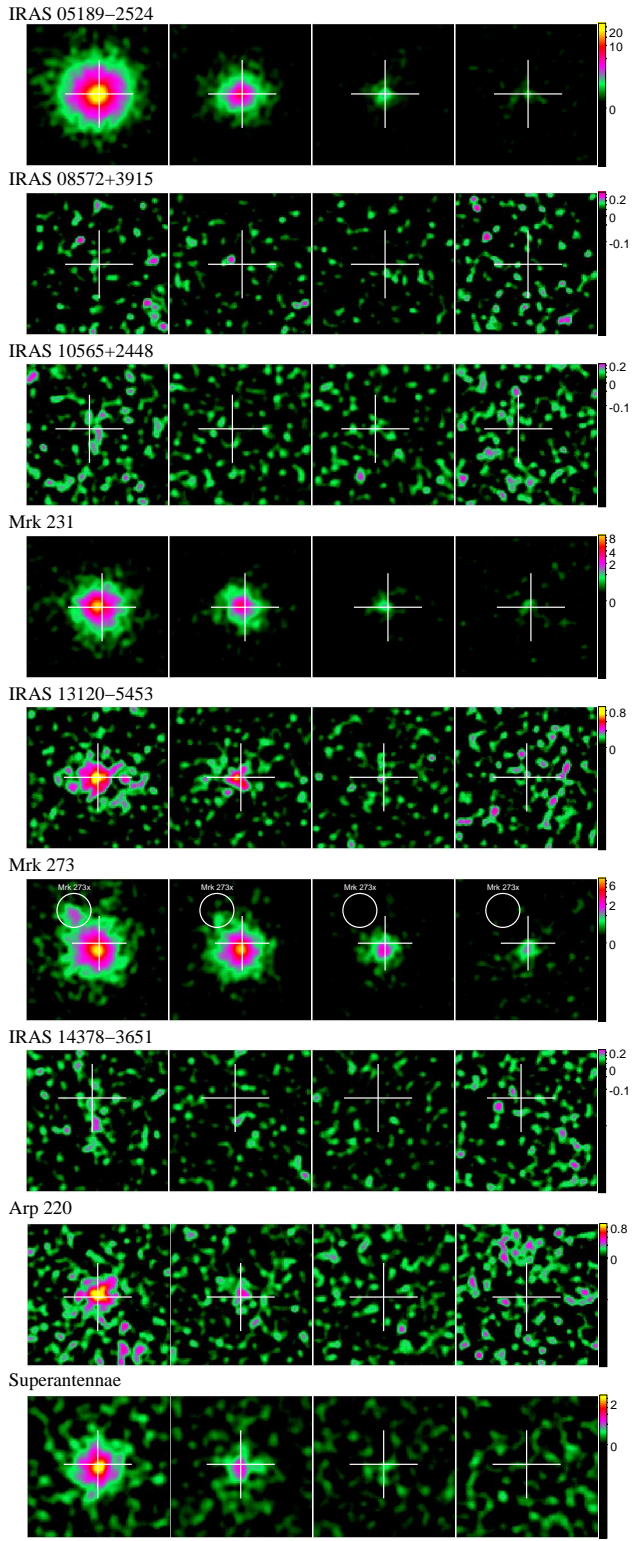}
\caption{{\it NuSTAR} background-subtracted images of each target in the
  (from left to right)
  3--10, 10--20, 20--30, and 30--79 keV energy bands.  After background
  subtraction, the FPMA and FPMB
  data were added together to improve the signal to noise.  In the
  cases where there are multiple exposures, all the exposures were
  also added together.  Each image is 2$\arcmin$ on a side and smoothed with 4-pixel
  Gaussians. A 2$\arcmin$ wide white cross is centered on the NED position of each
  target, with the exception of Mrk~273 whose cross is 1.6$\arcmin$.  The position of the background source, Mrk~273X, is identified by a circle in the images of Mrk~273.
  The color scale is log counts. }  
\label{fig:postage}
\end{figure}

\section{{\it NuSTAR} Photometric Results}
\label{sec:phot}

For each FPM, we created total, background, and background-subtracted images in four bands: 3--10, 10--20,
20--30, and 30--79 keV.  These images were produced using the 
{\tt  nuskybgd} code designed to simulate the total
background \citep{wik14}.  The {\tt nuskybkg} code takes into account
the telescope response and the energy dependence of the background
when producing the background images.  To improve photon statistics, we co-added 
the images from the two FPMs and all epochs if multiple observations
exist.  In Figure~\ref{fig:postage}, we show image stamps of our
co-added background-subtracted images.  Using the same circular regions
as the spectral extraction regions, we determined the total,
background, and background-subtracted (net) counts in our images.
These values are tabulated in Table~\ref{tab:photo}.  Two
of our targets (IRAS~08572+3915 and IRAS~10565+2448) were not
detected in any of the four bands and one of the targets (IRAS~14378--3651) 
was detected in only the 3--10~keV band.  Conservatively, we assume that sources are undetected in a
given band if the net counts are less than three times the estimated
error.  The error is calculated as:   $\sigma =
  \sqrt{N^2+B^2+\sigma^2_{\rm bkgsys}}$ where $N$ is the counting error on
  the number of net counts, $B$ is the counting error on the number of
background counts, and $\sigma_{\rm bkgsys}$ is the systematic error from
the background simulations.  $\sigma_{\rm bkgsys}$ is assumed to be 10\%
of the background counts below 20~keV and 3\% of background counts above 20~keV
(see \citealt{wik14} for more details.)

\begin{deluxetable*}{llllllllll}
\tablecolumns{10}
\tabletypesize{\tiny}
\setlength{\tabcolsep}{0.01in}
\tablecaption{{\it NuSTAR} Photometry Results\tablenotemark{a}}
\tablewidth{0pt}
\tablehead{\colhead{Source} & \colhead{3--10~keV}& \colhead{NBR}&\colhead{10--20~keV}& \colhead{NBR}& \colhead{20--30~keV}
& \colhead{NBR}&\colhead{30--79~keV}&
\colhead{NBR}&\colhead{$\Gamma_{\rm eff}$}\\
\colhead{(1)}&\colhead{(2)}&\colhead{(3)}&\colhead{(4)}&\colhead{(5)}&\colhead{(6)}&\colhead{(7)}&\colhead{(8)}&\colhead{(9)}&\colhead{(10)}
}
\startdata
IRAS 05189--2524&8348.6$\pm$94.5&15.7&1870.1$\pm$46.3&7.4&360.9$\pm$24.3&1.6&169.7$\pm$27.4&0.3&1.9\\
IRAS 08572+3915&(14.7$\pm$12.1)&\nodata&(5.5$\pm$9.2)&\nodata&(0.1$\pm$9.4)&\nodata&(--15.1$\pm$15.6)&\nodata&\nodata\\
IRAS 10565+2448&(17.7$\pm$14.2)&\nodata&(0.4$\pm$9.2)&\nodata&(0.7$\pm$9.4)&\nodata&(--10.7$\pm$15.7)&\nodata&\nodata\\
Mrk 231&2251.7$\pm$50.8&7.6&993.1$\pm$34.3&5.9&223.7$\pm$20.9&1.1&158.9$\pm$27.7&0.3&1.3\\
IRAS 13120--5453&234.8$\pm$19.7&1.7&107.1$\pm$13.8&1.4&(30.9$\pm$10.9)&\nodata&(24.1$\pm$16.5)&\nodata&1.3\\
Mrk 273&2102.7$\pm$50.2&5.6&1610.5$\pm$42.2&10.4&473.6$\pm$24.8&3.4&222.5$\pm$24.8&0.6&0.7\\
IRAS 14378--3651&53.2$\pm$14.1&0.4&(10.1$\pm$9.6)&\nodata&(--8.4$\pm$9.7)&\nodata&(5.1$\pm$15.4)&\nodata&\nodata\\
Arp 220&265.0$\pm$24.9&0.8&69.8$\pm$16.5&0.4&(7.5$\pm$14.9)&\nodata&(1.7$\pm$24.8)&\nodata&1.8\\
Superantennae&762.5$\pm$35.8&1.6&223.0$\pm$22.4&0.9&(38.6$\pm$17.9)&\nodata&(--0.8$\pm$28.1)&\nodata&1.7\\
\enddata
\tablenotetext{a}{Values in parentheses indicate non-detections, but
  are included here for completeness.  Since the backgrounds are simulated, we conservatively assume that
  sources are undetected in a given band if the net counts are less
  than 3 times the error.  The errors are calculated as: $\sigma =
  \sqrt{N^2+B^2+\sigma^2_{\rm bkgsys}}$ where $N$ is the counting error on
  the number of net counts derived by subtracting the simulated background counts from the detected source counts in the same extraction region, $B$ is the counting error on the number of simulated
background counts, and $\sigma_{\rm bkgsys}$ is the systematic error from
the background simulations.  $\sigma_{\rm bkgsys}$ is assumed to be 10\%
of the background counts
at below 20~keV and 3\% of background counts at above 20~keV \citep[see][for more details]{wik14}.}
\tablecomments{
Col.(1): Source name.  Col.(2):  Net counts in the 3--10~keV band.
Col.(3): Net-to-background counts ratio in the 3--10~keV band.
Col.(4): Net counts in the 10--20~keV band.  Col.(5):
Net-to-background counts ratio in the 10--20~keV band.  Col.(6): Net
counts in the 20--30~keV band.  Col.(7): Net-to-background counts
ratio in the 20--30~keV band.  Col.(8): Net counts in the 30--79~keV
band.  Col.(9): Net-to-background counts ratio in the 30--79~keV
band.  Col.(10): Effective photon index calculated using the 3--10 and
10--20~keV counts ratio by assuming a simple power law.
}
\label{tab:photo}
\end{deluxetable*}

Based on the co-added images, the three brightest of our targets were detected 
at energies above 30 keV:  IRAS~05189--2524,  Mrk 273, and Mrk 231.
Six of our nine targets
 were detected in both the 3--10 and the 10--20 keV bands.  
Assuming a simple power law continuum, we calculated
an effective photon index ($\Gamma_{\rm eff}$) using the nominal count
ratio between the 3--10 and 10--20 keV bands.  These effective photon
indices are also listed in Table~\ref{tab:photo}.  In particular, the
estimated photon index for Mrk~231 is $\sim$1.3, approximately consistent
with the $\sim$1.4 derived from complex spectral fitting
\citep{teng14}.  Also of note is the estimated photon index of
Mrk~273.  With $\Gamma_{\rm eff} \sim$0.7, Mrk~273 is the only source with a hard spectrum among our detected targets,
perhaps implying heavy obscuration.

\section{Broadband X-ray Spectroscopic Results}
\label{sec:spec}  
The modeling of the broadband (0.5--30~keV) X-ray spectrum of Mrk~231 revealed a
surprising result:  the AGN in Mrk~231 appears to be intrinsically X-ray weak
and Compton-thin
rather than Compton-thick \citep{teng14}.  Following the success of
the Mrk~231 results, we fit the contemporaneous broadband
spectrum for the deep survey sources.  For IRAS~13120--5453, the only
shallow survey object with a {\it NuSTAR} spectrum, we use archival {\it
  Chandra} data to anchor the low energy portion of the spectrum.

\citet{teng14} demonstrated the importance of constraining both the starburst
and AGN contributions to the X-ray spectrum.  
To model the starburst contribution, we include both the
thermal and non-thermal components.  The thermal component is
represented by one or two MEKAL components and the non-thermal
component is a cutoff power law with $\Gamma$ fixed at 1.1 and a
cutoff energy of 10~keV.  Unless stated otherwise in the text for each source,
the normalizations of these two
components are held fixed so that their luminosities are consistent
with the \citet{mineo12a, mineo12b} relations based on the target's
star formation rate (SFR).  For the AGN contribution, we use power
law based models to estimate the AGN luminosity.  These include an
absorbed power law, the AGN torus models MYTorus \citep{mytorus} 
and BNTorus \citep{bntorus},   and reflection models.  
When appropriate, we also include Gaussian emission lines.
The best-fit model parameters and results for the Deep Survey ULIRGs
are tabulated in Table~\ref{tab:fits}.

\subsection{IRAS 05189--2524}
\label{sec:spec05189}

\begin{figure}[h]
\centering
\includegraphics[width=2.5in, angle=270]{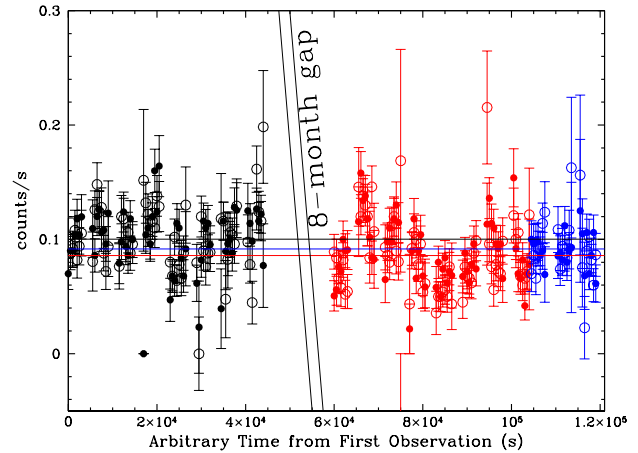}
\caption{Background-subtracted {\it NuSTAR} 3--10~keV light curve of IRAS~05189--2524.  The solid
  points are data from FPMA and the open points are data from FPMB.
  Each data point represents a 500 s temporal bin.  The data from the
  first, second, and third observations are represented in black, red,
and blue, respectively.  The colored lines show the average count rate
for each observation.  There is a $\sim$20\% decrease in the average count
rate from the first to the second pointing.}  
\label{fig:f05189var}
\end{figure}

IRAS~05189--2524 is the X-ray brightest ULIRG in our sample.
Observed in two {\it NuSTAR} pointings separated by about eight months, the second of which was
divided into two data sets, IRAS~05189--2524 has shown minor
variability between these epochs.  In Figure~\ref{fig:f05189var}, we show
the light curve for our {\it NuSTAR} observations.  The
average count rate in the second and third observation changed by
$\sim$20\% relative to the average count rate ($\sim$0.1 counts per second) in the first
observation.  However, we note that this variation is smaller than the standard deviation ($\sim$0.03 counts per second or $\sim$30\%) derived when data points from all three observations are combined.
Therefore, the variability is not statistically significant.
Emission lines at 6.4 and 6.8~keV were detected, with a $\Delta \chi^2$ of 65 for four degrees of freedom.

Although the light curve shows apparent minor variability, there is no obvious
variation in the spectral shape between {\it NuSTAR} observations.
 In particular, we do not see the drop in the
2--10~keV emission measured by \citet{teng09} from their {\it Suzaku} data;
the source appears to have reverted back to its previous ``high'' state.  
Therefore, we combined the spectra from all
three {\it NuSTAR} data sets for our broadband modeling.

\begin{figure}[h]
\centering
\includegraphics[width=2in, angle=270]{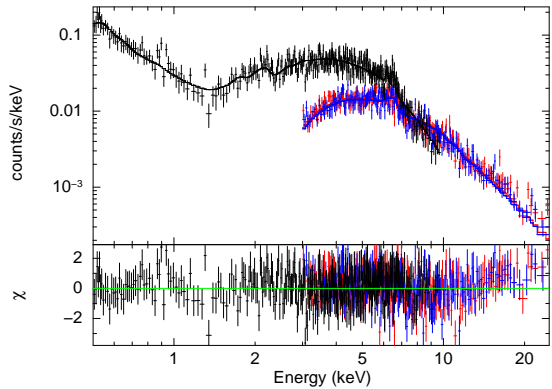}
\caption{Best-fit model of two partial covering absorbers displayed
  with the IRAS~05189--2524 data.  The {\it XMM-Newton}
  data are displayed in black (binned to at least 25 counts per bin) while the co-added {\it NuSTAR} data
  are displayed in red (FPMA) and blue (FPMB; binned to 4 sigma).  The modeling implies 
  that IRAS~05189--2524 hosts a luminous Compton-thin AGN.}  
\label{fig:f05189}
\end{figure}

Modeling multiple epochs of historic X-ray data, \citet{teng09}
found that the best-fit model to explain the sudden change in spectral
shape is an increase in the line-of-sight column from two partial
covering absorbers.  Following this result, we fit the {\it
  XMM-Newton} plus {\it NuSTAR} broadband spectrum with a double
partial covering model for the AGN component.  Following the best-fit
model for Mrk~231 in \citet{teng14}, we also included a MEKAL and a
cutoff power law ($\Gamma = 1.1$ with cutoff energy at 10~keV) for the
non-thermal emission from high--mass X-ray binaries (HMXBs) to account for the $\sim$80~$M_\odot$~yr$^{-1}$ starburst, estimated from the infrared luminosity.
 The photon index for the AGN, $\Gamma=2.51\pm0.02$, 
is very steep,  but is not far from the top of the range \citep[e.g., $1.5<\Gamma<2.2$;][]{nandra, reeves}
observed in other AGN.  The two partial covering
absorbers have $N_H$ of $5.2\pm0.2 \times 10^{22}$~cm$^{-2}$ and
$9.3^{+1.0}_{-0.7} \times 10^{22}$~cm$^{-2}$ with 98$\pm0.2$\% and
74$^{+1.2}_{-1.6}$\% covering fractions, respectively.  This model is shown ($\chi^2_\nu \sim$1.07), along
with the spectrum, in Figure~\ref{fig:f05189}.  The derived $\Gamma$ is steeper than $\Gamma_{\rm eff}$ estimated from the photometry because the $\Gamma_{\rm eff}$ calculation did not account for the flat power law contribution from the HMXBs ($\Gamma_{\rm HMXB} = 1.1$).  

Given the high 2-10~keV flux levels, we do not expect this source
to be highly obscured.  Both the MYTorus and BNTorus models
poorly  describe the broadband spectrum ($\chi^2_\nu \sim$1.8  for both).

From our modeling, we find that the AGN in IRAS~05189--2524 is
currently in a Compton-thin state.  
Its intrinsic 2--10~keV luminosity is $3.7\times
10^{43}$~erg~s$^{-1}$, about 0.8\% of the bolometric luminosity.
Our present result is consistent with those from the multiple-epoch
fitting by \citet{teng09}.  The large drop in the observed 2--10~keV
flux in the 2006 {\it Suzaku} XIS data is likely due to an intervening
absorber, since a factor of 30 change in intrinsic flux is rare in
AGN \citep{Gibson12}.  
The incomplete nature of the time series data makes it impossible to 
precisely determine the timescale of variability.
If it is on the order of years, as is consistent with the data, 
then 
the Compton-thick absorber responsible for the {\it Suzaku} variability 
must be within a few pc of the nucleus.   
However, a more extreme case of intrinsic variability (a factor
of $\sim$260) was observed in the narrow-line Seyfert 1 galaxy PHL~1092
\citep{miniutti12} so we cannot absolutely rule out the possibility of strongly varying intrinsic flux. 
Indeed, our best-fit model, even in the high state, should not have been detectable by
{\it Suzaku} PIN in the 2006 observation, which speaks to the much greater sensitivity 
above 10 keV of {\it NuSTAR} compared to {\it Suzaku} PIN.

\subsection{IRAS 13120--5453}
\label{sec:spec13120}

\begin{figure}[h]
\centering
\includegraphics[width=2in, angle=270]{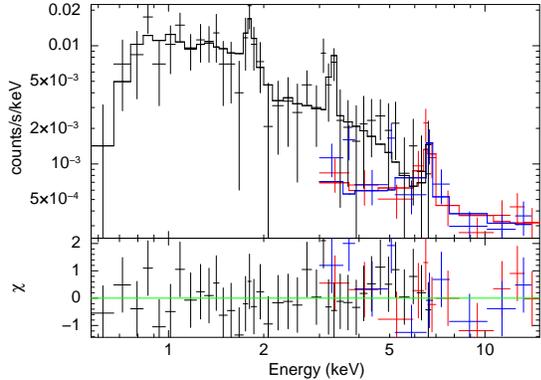}
\caption{Best-fit model with MYTorus components to the IRAS~13120--5453 data.  The {\it Chandra}
  data from 2006 binned to at least 15 counts per bin are displayed in black while the 2014 {\it NuSTAR} data
  binned to 3 sigma are displayed in red (FPMA) and blue (FPMB).  The modeling is
  consistent with IRAS~13120--5453 hosting a Compton-thick AGN.}  
\label{fig:f13120}
\end{figure}

We first fit the broadband spectrum of IRAS~13120--5453 with a simple power law modified
only by Galactic absorption and a MEKAL model for the starburst
component.  Because of the relatively high Galactic column density ($N_{\rm H, Gal} \sim$4.6$\times10^{21}$~cm$^{-2}$),
it is difficult for the MEKAL model to constrain the line components
below 2~keV.  The best-fit power law requires $\Gamma$ $\sim -2$
to fit the shape of the $>2$~keV spectrum, implying a 
highly obscured AGN continuum.  Additionally, the spectrum shows three
strong emission lines at 1.86, 3.40, and 6.78 keV which correspond to
Si~XIII, Ar~XVIII, and Fe~XXV, respectively.  
The $\Delta \chi^2$ for the iron line is 8.35 for two degrees of freedom.
The strong lines of the
$\alpha$-elements Si (EW$\sim$0.20~keV) and Ar (EW$\sim$0.53~keV) may be an indication of a strong
starburst.  Since the lines are  stronger than the predictions of the MEKAL plasma model, 
if they are starburst in origin then a more complex starburst model is needed, for example
with multiple temperatures, or highly non-solar abundances.

To constrain the intrinsic absorption of the AGN continuum, we added
an absorption component and fixed $\Gamma$ at the canonical value of
1.8.  This new model requires that the intrinsic absorber have a
column density of $\sim 4\times10^{24}$~cm$^{-2}$, implying the AGN is
Compton-thick.  For both torus  models, we fixed
the torus inclination angle at 85$^\circ$, since 
the optical data suggest the AGN is Type~2.  
Since we do not have an independent measure of the star formation rate in
IRAS~13120--5453 as we did for Mrk~231 and IRAS~05189--2524, we allowed the hot gas
temperature and the normalizations of the HMXB model components to
vary.  We also included Gaussian components to model the lines
at 1.9, 3.4, and 6.8 keV.  

The MYTorus model seems to fit the
spectrum well ($\chi^2_\nu \sim 0.80$); however, the model cannot
constrain the error of $\Gamma$ within the bounds of the MYTorus model
($1.4 < \Gamma < 2.5$).  For the best-fit model, we fixed $\Gamma$ at 1.8
resulting in a column density of 3.1$^{+1.2}_{-1.3} \times
10^{24}$~cm$^{-2}$, consistent with the assertion by \citet{iwasawa11}
that IRAS~13120--5453 is Compton-thick based on the strength of the Fe
line.  The 2--10~keV absorption-corrected luminosity for
IRAS~13120--5453 is $1.25\times 10^{43}$~erg~s$^{-1}$.  The
starburst component is absorbed by a column of 5.6$^{+14.3}_{-5.6}
\times 10^{21}$~cm$^{-2}$.  The thermal component has a temperature of
0.56$^{+0.16}_{-0.31}$~keV and a 0.5--2~keV luminosity of 1.8$\times
10^{41}$~erg~s$^{-1}$.  The non-thermal HMXB component has a
0.5--8~keV luminosity of 4.5$\times 10^{41}$~erg~s$^{-1}$.  The
luminosities of both the
thermal and non-thermal components are consistent with a star
formation rate of $\sim$170~M$_\odot$~yr$^{-1}$, based on the \citet{mineo12a, mineo12b} relations.  Although high, this star
formation rate is within the range observed for 
ULIRGs.  For comparison, Mrk~231 has a star formation rate of
$\sim$140~M$_\odot$~yr$^{-1}$ \citep{rv11}.  
The absorption-corrected 2--10~keV luminosity of IRAS~13120--5454 is $\sim$0.67\% of its
AGN bolometric luminosity.  The best-fit MYTorus spectrum is shown in Figure~\ref{fig:f13120}.
Given the data quality, the BNTorus model places poor constraints on the torus opening angle
and $\Gamma$.  However, the values for the column density, intrinsic X-ray luminosity,  and the
starburst components are consistent with the results from MYTorus.

\subsection{Mrk 273}
\label{sec:specmrk273}

Mrk~273 is the only source in a past {\it Suzaku} survey of ULIRGs that
was detected above 10~keV \citep{teng09} by PIN.  With only a marginal detection
(1.8$\sigma$), the {\it Suzaku} data
required a double partial covering model and implied that at least one
of the partial covering absorbers is Compton-thick ($N_H \sim 1.6\times10^{24}$~cm$^{-2}$).  
A variability
analysis using the {\it Suzaku} and historic X-ray data 
suggests that the variations in spectral shape between 2--10~keV
are due to changes in the column density or the fraction of the partial
covering.

\begin{figure}[h]
\centering
\includegraphics[width=2in, angle=270]{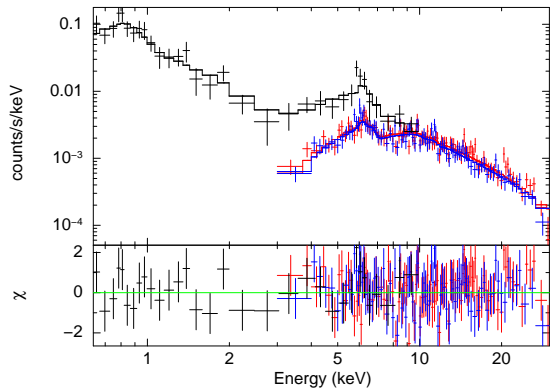}
\caption{The Mrk~273 broadband data modeled using the MYTorus model.
  The {\it XMM-Newton} EPIC-pn data are shown in black (binned to at
  least 15 counts per bin) while
  the simultaneous {\it NuSTAR} data are shown in red (FPMA) and blue
  (FPMB).  The source is well-detected by {\it NuSTAR}, the data from
  which were binned to 4$\sigma$.}  
\label{fig:mrk273}
\end{figure}

Mrk~273 is well-detected by {\it NuSTAR} up to
$\sim$30~keV, above which the background dominates. 
 Both the MYTorus and BNTorus  models fit the data
nearly equally well.  As with the Mrk~231 analysis \citep{teng14}, we
included model components that account for HMXB and thermal emission
for a starburst that is forming stars at a rate of
$\sim$160~$M_\odot$~yr$^{-1}$ \citep{vei09a}.  With the torus
inclination angle fixed at 85$^\circ$, 
the best-fit MYTorus model ($\chi^2_\nu = 0.91$) suggests that the 
direct intrinsic AGN emission (or the zeroth order emission)
has $\Gamma = 1.43^{+0.17}_{-0.03}$, and the global $N_H$ is
$4.4 \pm 0.1 \times10^{23}$~cm$^{-2}$.  The leaky fraction of the absorber is
only 3.1$^{+2.4}_{-1.8}$\%.  This best-fit model and the data are
shown in Figure~\ref{fig:mrk273}.

The BNTorus model gives similar results ($\chi^2_\nu=0.90$).
With the torus inclination angle fixed at 87$^\circ$, $\Gamma$ is
constrained to be 1.29$^{+0.20}_{-0.17}$ and the line-of-sight $N_H$ is
$3.5^{+1.1}_{-0.8}\times10^{23}$~cm$^{-2}$.  The opening angle of the
torus is at least $\sim$43$^\circ$.  The BNTorus model infers a leaky
fraction of 4.5$^{+4.3}_{-2.7}$\%.  These parameters are consistent
with those derived using the MYTorus model.

\citet{teng09} concluded that the spectral variability seen in Mrk~273
was due to changes in the column density.  By fitting the multiple
epoch data together and assuming a common model, they showed
that the older measurements of the column density with {\it Chandra} and
{\it XMM-Newton} were a factor of 2--4 lower than the
{\it Suzaku} measurement.  Our {\it NuSTAR} analysis is consistent
with this result, as the column density we derived using the torus
models is a factor of $\sim$4 lower than the {\it Suzaku}
measurement.  Similar to our Mrk~231 results \citep{teng14}, the torus
models favor a relatively flat power law intrinsic photon index for an AGN.  The intrinsic
2--10~keV luminosity of Mrk~273 is $\sim$8.6$\times
10^{42}$~erg~s$^{-1}$, representing $\sim$0.3\% of the AGN bolometric luminosity.

\subsection{Arp 220}
\label{sec:specarp220}

Although it is the nearest ULIRG, Arp~220 is the faintest source at {\it NuSTAR} energies
in our Deep Survey.  This source was observed by {\it
  Suzaku} in 2006, but was undetected above 10~keV \citep{teng09}.
The {\it Suzaku} spectrum, with a lack of detection above 10~keV and
an ionized Fe line with large EW ($\sim$0.42~keV), suggests that
the AGN is very heavily obscured.  
The direct emission from the AGN is completely
obscured by a high column density, leaving behind a purely reflected
spectrum.  Since the {\it NuSTAR} flux between
15--40~keV is 35 times lower than the upper limit derived from the
{\it Suzaku} observations, our {\it NuSTAR} observations can more tightly
constrain the X-ray properties of Arp~220.

When modeling the new broadband X-ray spectrum, we first revisited the ionized 
reflection model favored by \citet{teng09}, which did not include an HMXB component. 
The MEKAL plus ionized reflection model \citep[reflionx;][]{reflion} is
well-fit to the broadband data ($\chi^2_\nu \sim$1.25).  All model parameters
are consistent with those derived from the {\it Suzaku} XIS data
alone.  To better model the shape of the spectrum below 2~keV, we
added a second MEKAL component. 
The best-fit hot
gas temperatures are $0.10^{+0.02}_{-0.10}$ and $0.50^{+0.20}_{-0.25}$~keV.  Both these
temperatures are consistent with those observed in ULIRGs
\citep[e.g.,][]{ptak03, frances03, teng05, teng09, teng10}.  The
underlying reflected power law has $\Gamma = 1.76^{+0.22}_{-0.32}$, assuming the
input ionization parameter $\xi$
is 10$^3$~erg~cm~s$^{-1}$.  The reflected 2--10~keV luminosity is
$\sim$9.0$\times 10^{40}$~erg~s$^{-1}$.  
As the detailed modeling of
Mrk~231 by \citet{teng14} has shown, it is important to constrain the
HMXB component that also contributes to the X-ray spectrum.  When an
HMXB component \citep[SFR=200~$M_\odot$~yr$^{-1}$  assuming the starburst infrared luminosity derived by][]{vei09a} is added to the model, the reflection component is no
longer required.

\begin{figure}[h]
\centering
\includegraphics[width=2in, angle=270]{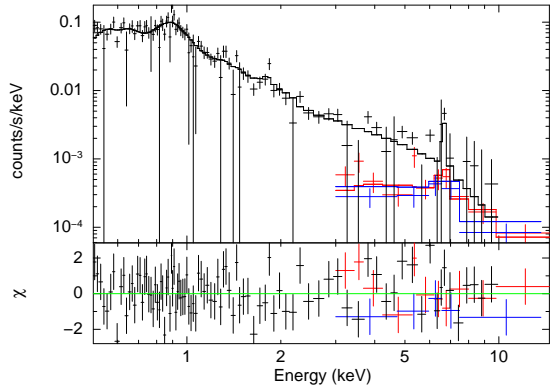}
\caption{The Arp~220 broadband data modeled without an AGN component.  The {\it XMM-Newton} EPIC-pn data, binned to at least 15
  counts per bin, are shown in black while
  the simultaneous {\it NuSTAR} data, binned to 3 sigma, are shown in red (FPMA) and blue
  (FPMB).  The FPMB data were of poorer quality than those of FPMA,
  likely due to a higher (simulated) local background.  It is possible
that, due to the high column density, the observed X-ray spectrum does
not show any signatures of an AGN.}  
\label{fig:arp220}
\end{figure}

We attempted to use the BNTorus model to constrain the properties of the
obscuring torus.  However, due to the poor photon statistics
above 10~keV, the model does a poor job in deriving robust values and
errors for parameters that characterize the torus.  The model suggests
a line-of-sight column density of at least 5.3$\times
10^{24}$~cm$^{-2}$ with the best-fit value tending toward $>
10^{25}$~cm$^{-2}$.  Similarly, the MYTorus model has difficulty
constraining the properties of the torus.  Using this model, the
column density is at least 1.2$\times 10^{24}$~cm$^{-2}$, with the
nominal value hitting the upper bound of the model at
$10^{25}$~cm$^{-2}$.  In this case, it is not possible to 
measure the intrinsic X-ray luminosity of the AGN.  
Therefore, if an AGN is present in Arp~220, it is highly Compton-thick and the $N_{\rm H}$ cannot be constrained with the {\it NuSTAR} $>$10~keV data.  
This result is
consistent with measurements by two groups that find that the 
western nucleus of Arp~220 is embedded in a column of 
$1.3\times 10^{25}$~cm$^{-2}$   \citep{downes07} 
to $1.5\times 10^{25}$~cm$^{-2}$ \citep{scoville14}.

It is possible that the observed X-ray spectrum of Arp~220 does not
have an AGN component at all.  If the column density is $>
10^{25}$~cm$^{-2}$, then no direct emission can be detected if the
absorber is not ``leaky''.  What the observed global X-ray spectrum
represents is simply the thermal and non-thermal emission from the
major starburst. 
Therefore, we also modeled the Arp~220 spectrum
without an AGN component.  
The data are well-fit by a two-temperature MEKAL plus a strong ionized Fe~K line.  
The iron line likely originates from the strong starburst.
A strong bremsstrahlung component would be expected to accompany the 6.7~keV emission.  Thus, we also included a redshifted bremsstrahlung component to the two-temperature MEKAL model.  The MEKAL luminosities were fixed such that they are consistent with the star formation rate based on the \citet{mineo12b} relation.  This best-fit model ($\chi^2_\nu=1.22$) is shown in Figure~\ref{fig:arp220}.  

To summarize, the X-ray emission from Arp~220 appears to be consistent with only a starburst.  
However, there is the possibility that a very deeply buried AGN is present in this source.  

\subsection{The Superantennae}
\label{sec:specsuper}

\citet{braito09} reported that the Superantennae was detected by {\it
  Suzaku} PIN above 10~keV.  Although \citet{braito09} claim a signal-to-noise ratio of $\sim$10 in their 15--30~keV detection, the source spectrum is only $\sim$5.5\% above 
the PIN background, which has a systematic uncertainty of $\sim$1.5\%.  They describe the source as a Compton-thick AGN ($N_H
\sim 3-4 \times 10^{24}$~cm$^{-2}$) shining with an intrinsic
2--10~keV luminosity of a few times 10$^{44}$~erg~s$^{-1}$, at the
level of a luminous quasar.  This galaxy was observed twice by {\it
  NuSTAR}, with a temporal separation of about four months.  In both
epochs, the Superantennae is weakly detected, with a 15--30~keV flux 
that is 30 times lower than that measured by  \citet{braito09} with {\it Suzaku}. 
There appears to be no significant variability between the two sets of
observations in terms of the spectral shape or the strengths of the Fe  
lines.  Due to the lack of discernable variability, we have co-added
the two sets of {\it NuSTAR} spectra in order to improve the signal-to-noise
ratio of the overall spectrum. 
Emission lines at 6.5 and 6.9~keV were detected with 
$\Delta \chi^2=30$  for four degrees of freedom.

\subsubsection{Broadband Fitting}

\begin{figure}[h]
\centering
\includegraphics[width=2in, angle=270]{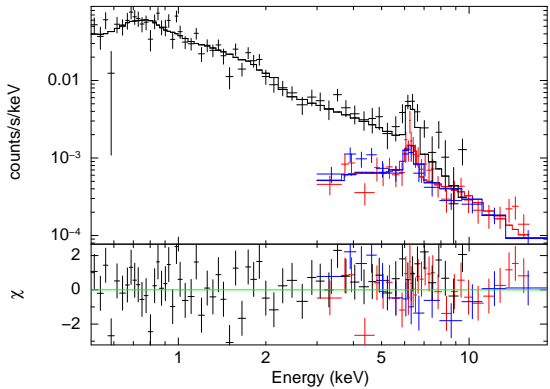}
\caption{The Superantennae broadband data modeled by a single power
  law AGN component as well as star formation components.  The {\it XMM-Newton} EPIC-pn data, binned to at
  least 50 counts per bin, are shown in black while
  the {\it NuSTAR} data, binned to 3 sigma, are shown in red (FPMA) and blue
  (FPMB).  The {\it NuSTAR} data shown are the co-added spectrum of
  two epochs separated by about four months.  }  
\label{fig:sa}
\end{figure}

We have applied three different models to the broadband spectrum of the Superantennae.  
All three include thermal and power law components for the starburst in
addition to the typical AGN component.  We first tested whether a
simple absorbed power law can explain the spectral shape.  A point in
favor of the Compton-thick AGN scenario is that the power law spectrum 
inferred from only $<$10~keV data is relatively flat \citep[$\Gamma \sim$1.3; ][]{braito03}.
With the broader energy coverage of the {\it XMM-Newton} plus {\it NuSTAR} data,
we find that the spectrum, after accounting for the starburst contribution, 
can be well-fit with a standard 
power law model for the AGN component ($\chi^2_\nu = 1.30$).  The
best-fit result requires only Galactic absorption and $\Gamma$ of 1.54$\pm$0.13, consistent with
that measured from typical AGN.  We do not detect the presence of a
strong neutral Fe line.  The apparent Fe emission can be
described by two narrow Gaussians with central energies at
6.54$^{+0.16}_{-0.07}$ and 6.87$^{+0.37}_{-0.10}$~keV.  These lines
have EWs of 288$^{+370}_{-94}$ and 296$^{+521}_{-163}$~eV,
respectively.  This fit is shown in Figure~\ref{fig:sa}.  

For completeness, we also applied the torus models to the spectrum to
test whether the AGN can also be Compton-thick as suggested by
\citet{braito09}.  Both the MYTorus and BNTorus models
give similar parameter values;  however, the
BNTorus models cannot constrain the opening angle of the torus. 
The MYTorus model fits the data very well ($\chi^2_\nu=1.27$).
The best-fit model implies that the underlying nuclear spectrum, with 
$\Gamma = 1.54^{+0.17}_{-0.14}$, is obscured by a column of $N_H =
4.2^{+5.8}_{-3.1}\times10^{24}$~cm$^{-2}$.  A nominally small fraction,
13$^{+87}_{-9}$\%, of the direct emission is leaked through the
absorber, but the large error bars clearly indicate a poorly constrained parameter.  

Both the Compton-thin and Compton-thick models are statistically
equivalent.  Although technically the Compton-thick model has a
smaller reduced $\chi^2$, it is a more complicated model that does a poor job of
constraining the leaked emission component.  If the leaked fraction parameter is left
completely free, the parameter errors reach an unphysical value.  
If the leaked component is
removed, MYTorus cannot account for most of the 3--9 keV
emission in the spectrum.  Therefore, we favor the Compton-thin
interpretation of the Superantennae spectrum.  With this model, the
intrinsic 2--10~keV luminosity of the AGN is
$1.7\times10^{42}$~erg~s$^{-1}$, several hundred times
lower than that reported by \citet{braito09}.  The 2--10~keV to
bolometric luminosity for this AGN is 0.08\%.

In these models, the
15--30~keV flux for the Superantennae is $\sim$1.7$\times10^{-13}$
erg~s$^{-1}$~cm$^{-2}$, a factor of 30 lower than the flux measured in
this band by \citet{braito09} with {\it Suzaku}.  There does not
appear to have been any notable variability below 10~keV for the
Superantennae.  
\citet{braito09} noted two other sources within the {\it Suzaku} PIN
non-imaging field-of-view that are also AGN and have similar fluxes to the
Superantennae.  While they used {\it XMM-Newton} data to constrain the
spectral properties of these background sources and concluded that
they can be described by unabsorbed power laws, it is possible that
their contributions above 10~keV were not fully accounted for.  
Other field sources could also have contaminated the {\it Suzaku} 
$>10$~keV measurement. These include a field source within the {\it NuSTAR} field of view in the first observation of the Superantennae.  The field source is a point source $\sim$8$\farcm$5 from the Superantennae (RA: 19:32:48.3, Dec: --72:33:52.0).  Although fainter than the Superantennae, the count rate of this source is 36\% and 45\% of those of the Superantennae in the 3--10~keV and 10--20~keV bands, respectively.  These numbers suggest that the field source can harden the apparent 15--30~keV {\it Suzaku} PIN spectrum of the Superantennae at $>$10~keV, leading to the previous conclusion that the source is Compton-thick. There may be other field sources within the {\it Suzaku} field-of-view that are outside of the {\it NuSTAR} field-of-view with similar properties.  Therefore, we conclude that the {\it Suzaku} data were contaminated and that the Superantennae most likely 
hosts a Compton-thin AGN.  

\subsubsection{Iron Line Variability}

\citet{jia12} studied the Fe emission lines between 6 and 7 keV in the
Superantennae in X-ray observations spanning 8 years (2001--2009).
They found that the 6.4, 6.7, and 6.9 keV lines varied in the three
observations taken in this time range.  In 2001, {\it XMM-Newton}
detected emission lines with central energies consistent with the 6.4
and 6.9 keV lines, but in subsequent {\it Suzaku} XIS (2006) and {\it
  Chandra} (2009) observations, only the 6.7 keV line was
significantly
detected.  Furthermore, the EW of the 6.7 keV line varied over the
three observations.  

In our broadband fitting above, we found two narrow Gaussians
with energies consistent with the 6.7 and 6.9 keV lines.  These lines
have approximately the same EW.  We examined the Fe line complex in
our 2013 {\it XMM-Newton} data in more detail.  By
using the unbinned spectrum with the Cash statistic in XSPEC, we see
signatures of the 6.7 and 6.9 keV lines.  There is also a hint of 
the 6.4 keV emission line that is not statistically significant.

We then compared the 2013 data with those taken by {\it XMM-Newton} in
2001.  We did not use the {\it Suzaku} or the {\it Chandra} data in
this study since those two telescopes have different responses in the
relevant energy range and we want to limit the variables in our
comparison.  Using the most up-to-date calibration, we re-analyzed the
2001 data in the same manner as the 2013 spectrum.  Although the 2001
data were shallower, it is clear that the 6.4 keV line was stronger in 2001
than in 2013 (using the same continuum model as for 2013).  The 6.7 keV line has
appeared since 2001 and the 6.9 keV line has grown stronger in 2013.
Therefore, the relative strengths of these lines have changed between 2001 and
2013.  It is unclear what caused the iron line EWs to change and the 
6.4~keV line to disappear within a decade.  
One possibility is a change in the ionization state of the
accretion disk from which the line emission arises.  The variability of
the iron lines supports the conclusions of
\citet{jia12} that the lines must come from a compact region like the
central engine.

\subsection{The Undetected Sources: IRAS 08572+3915, IRAS 10565+2448,
  IRAS 14378--3651}
\label{sec:specundet}

IRAS 08572$+$3915, IRAS 10565$+$2448, and IRAS 14378$-$3651 were undetected
by {\it NuSTAR} in 25~ks.  Given the {\it NuSTAR} sensitivity limits,
this implies that the intrinsic 2--10 keV luminosities of these sources
are below $\sim 5\times 10^{42}$~erg~s$^{-1}$ for the typical
redshifts of our sources of $\sim$0.05, assuming
the standard canonical AGN power law model.  Otherwise, the strong
X-ray continua of these sources would have been detected above 10~keV. 

Using the observed count rates from our {\it NuSTAR} observations at
the locations of our targets, we determined upper limits to the
observed 2--10~keV luminosities of these sources, under the assumption that 
the obscuring column density is not high. 
 In the derivation,
we assumed only a power law component with $\Gamma = 1.8$ and Galactic
absorption.  {\bf  No additional column density was assumed.}
  Using the 3--10~keV count rates for each source, we used
WebPIMMS\footnote{https://heasarc.gsfc.nasa.gov/cgi-bin/Tools/w3pimms/w3pimms.pl}
to estimate the unabsorbed 2--10~keV luminosity. 
The total 3--10 keV count rates extracted from circular regions with 1\arcmin\ radii for 
IRAS 08572$+$3915, IRAS 10565$+$2448, and IRAS 14378$-$3651 are 
$3 \times 10^{-3}$, $2 \times 10^{-3}$, and $3 \times 10^{-3}$ counts per second, respectively.
Assuming no intrinsic obscuration, these rates correspond to upper limits to the 2-–10 keV luminosity 
of $\sim 6 \times 10^{41}$ and  $7 \times 10^{41}$~erg~s$^{-1}$ for IRAS 08572+3915 
and IRAS 10565+2448, respectively, and intrinsic 2-10 keV luminosity of 
$1 \times 10^{42}$~erg~s$^{-1}$ for IRAS 14378-3651.
These limits and measurement are consistent with those 
previously measured by {\it Chandra} \citep[e.g.,][]{teng09, teng10,
  iwasawa11}, and suggest that the intrinsic 2--10~keV
to bolometric luminosity ratios in these sources are 
$\lesssim$0.01,  $\lesssim$0.09, and 0.08\%, respectively, assuming no source obscuration.  
However, as we saw in the case of Arp~220 (\S\ref{sec:specarp220}), it is not possible to rule out
very high obscuring column densities.

\section{The Contribution of AGN Power to ULIRGs}
\label{sec:dis}

\subsection{The Hard X-ray Perspective}
\label{sec:hardxray}

For several decades, what powers the
enormous infrared luminosities of ULIRGs has remained an 
unanswered question.
Many studies concluded that the lack of strong X-ray detections in
ULIRGs implied that these sources are highly obscured
\citep[e.g.,][]{ptak03, frances03, teng05, iwasawa11}.  However, these
studies lacked sensitive detections at energies above 10~keV, which are
necessary to disentangle the effects of obscuration in order to robustly
measure the intrinsic X-ray luminosities of the AGN in ULIRGs.  With
the launch of {\it NuSTAR}, which is $\sim$100 times more sensitive
than {\it Suzaku} PIN in the 10--40~keV energy band, broadband
(0.5--30~keV) X-ray
spectroscopy has allowed us to estimate the intrinsic X-ray luminosity
of five of the nine ULIRGs in our sample and place constraints on the
remaining four.  

Our observations reveal that the ULIRGs in our sample have
surprisingly low observed fluxes in high energy ($>$10~keV)
X-rays.  Of the nine ULIRGs in our {\it NuSTAR} sample, six were
detected well enough to enable detailed spectral modeling of their
broadband X-ray spectra.  Of
these six, only one, IRAS~13120--5453, has a spectrum consistent with
a Compton-thick AGN. We cannot rule out the possibility that a second ULIRG 
in the sample, Arp~220, is highly Compton-thick ($N_H >10^{25}$~cm$^{-2}$).  
Thus, by the strictest definition 
($N_H >1.5 \times 10^{24}$~cm$^{-2}$),  these {\it NuSTAR} data show that most of 
the ULIRGs in our sample are {\bf not} Compton thick.

However, detailed analysis of the {\it NuSTAR} data on IRAS~05189--2524 and Mrk~273 shows that the hard X-ray fluxes of these sources have varied compared to similar {\it Suzaku} observations in 2006 \citep{teng09}.  The observed variability in both sources can be explained by a change in the absorbing column.  The column densities for both IRAS~05189--2524 and Mrk~273 have reduced by a factor of a few since the 2006 observations. These changes are sufficient to alter 
the classification of these AGN from being Compton-thick to Compton-thin.

In our sample of nine ULIRGs, three (IRAS~05189--2524, Mrk~231, and
Mrk~273) have strong hard X-ray continua above 10~keV after correcting for obscuration.  The remaining
targets have low count rates.  Only two ULIRGs (IRAS~05189--2524 and
IRAS~13120--5453) have intrinsic 2--10~keV AGN luminosities above
$10^{43}$~erg~s$^{-1}$.  Although the AGN in many of these ULIRGs dominate the spectral energy distributions at other wavelengths \citep[e.g.,][]{vei09a}, these active black holes do not appear to produce as 
much X-ray emission as would be expected for typical AGN.  When compared to their
bolometric luminosities, the AGN in our sample of ULIRGs are not emitting as much
X-ray power as Seyfert galaxies.  The intrinsic 2--10~keV to bolometric
luminosity is in the range of 0.03\% to 0.81\%.  For comparison, this value
ranges from 2 to 15\% for radio quiet quasars and Seyfert
galaxies \citep{elvis}. However, there is evidence that objects with Eddington ratios near or above unity have smaller 2--10~keV to bolometric
luminosity ratios \citep[0.3--0.7\%;][]{lusso10, lusso12, vf09}.  The
low 2--10~keV to bolometric luminosity ratios for our sample could
imply that these sources have high accretion rates.   This is unsurprising,
as ULIRGs are mergers that may be rapidly growing their central black holes.  
The high accretion rates of ULIRGs are consistent with the detection of ionized Fe~K lines in five 
of the six ULIRGs with spectra in our sample 
(IRAS~05189--2524, IRAS~13120--5453, Mrk~231, Arp~220, and the Superantennae). 
\citet{iwasawa12} suggested that there is a link between the presence of ionized lines in the COSMOS sample and high accretion rate.  
The exception is Mrk~273, which has a low accretion rate (see below).  

\begin{deluxetable}{llllll}   
\tablecolumns{6}
\setlength{\tabcolsep}{0.01in}
\tablecaption{Eddington luminosities and ratios}
\tablewidth{0pt}
\tablehead{\colhead{ULIRG} & \colhead{$M_{bh}$} & \colhead{$L_{Edd}$} & \colhead{$\lambda_{Edd}$} & \colhead{$l_x$} & \colhead{$\Gamma$}\\
\colhead{} & \colhead{(M$_\odot$)} & \colhead{(erg~s$^{-1}$)}& \colhead{} & \colhead{} & \colhead{}
}
\startdata
IRAS~05189--2524 & $3\times10^{7}$ & $3.6\times10^{45}$ &  1.2  & $1\times10^{-2}$ & 2.51\\
Mrk~231          & $2\times10^{7}$ & $2.1\times10^{45}$ &  5.2  & $2\times10^{-3}$ & 1.39\\
Mrk~273          & $6\times10^{8}$ & $6.9\times10^{46}$ &  0.04 & $1\times10^{-4}$ & 1.43\\
\enddata
\tablecomments{Columns:  
(1) ULIRG with measured dynamical black hole mass and spectrally
derived X-ray AGN luminosity; 
(2) Black hole mass; 
(3) Eddington luminosity; 
(4) Eddington ratio, $\lambda_{Edd}$=$L_{bol,AGN}$/$L_{Edd}$, where $L_{bol,AGN}$ is 
   estimated by multiplying column (3) by column (6) in Table~\ref{tab:sample}; 
(5) dimensionless 2--10~keV to Eddington luminosity ratio,  $l_x$=$L_X$/$L_{Edd}$;
(6) $\Gamma_{AGN}$, copied for convenience from Table~\ref{tab:fits} or \citet{teng14}.
}
\label{tab:edd}
\end{deluxetable}

For the three sources in our sample that have both dynamically measured
black hole masses \citep[e.g.,][]{vei09b} and spectrally derived AGN
fluxes, we calculated Eddington luminosities, Eddington ratios, and
2--10~keV to Eddington luminosity ratios, tabulated in Table~\ref{tab:edd}. 
These parameters distinguish the brightest objects in our sample each
into an unique category.  IRAS~05189--2524 is the X-ray brightest and
accreting at a super-Eddington rate.  
Mrk~231 is the X-ray faintest and also accreting at a super-Eddington rate. 
Mrk~273 has the most massive black hole and the lowest Eddington ratio.
Many studies have found that
$\Gamma$ becomes softer with increasing $\lambda_{Edd}$
\citep[e.g.,][]{shemmer05, shemmer08, brightman13}.  Comparing the
model-derived $\Gamma$ of our three brightest sources with their
$\lambda_{Edd}$ values, this $\Gamma$-$\lambda_{Edd}$ correlation
appears to hold with the exception of Mrk~231.  More recently,
\citet{yang14} found a correlation between $\Gamma$ and 
the dimensionless ratio of  2--10~keV and Eddington luminosities, which 
holds for  black hole accretion systems including both black hole binaries and AGN.  The
authors found that $\Gamma$ decreases with increasing $l_X$ up to
$\sim$10$^{-3}$ but steepens again above that $l_X$ value. 
(See also \citealt{Constantin09}.)  Our
limited data points from three sources are consistent with this
phenomenological model.  If both the $\Gamma$-$\lambda_{Edd}$ and
$\Gamma$-$l_X$ relations are true and indicative of the accretion
processes in most AGN, then Mrk~231 is an outlier.  This implies that
the 2--10~keV to bolometric luminosity correction is different for
Mrk~231 than for most AGN, as suggested by \citet{teng14}.

\begin{figure}[h]
\centering
\includegraphics[width=3in]{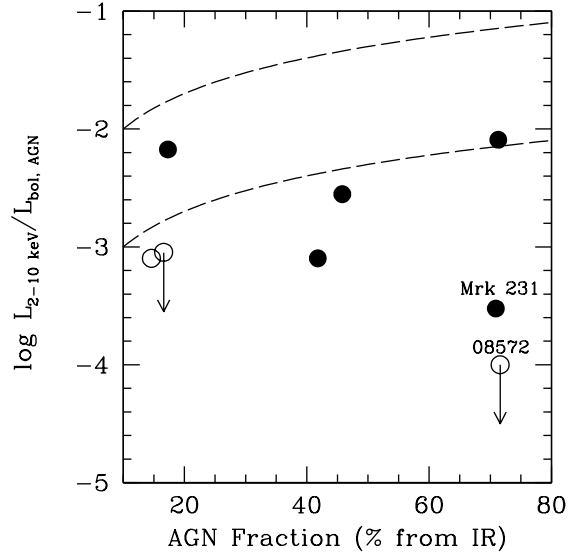}
\caption{Comparison of the intrinsic 2--10~keV to bolometric AGN luminosity ratio to
  the {\it Spitzer}-derived AGN fraction.  The AGN fraction represents
the fraction of the total infrared luminosity that is powered by the
AGN based on the methods of \citet{vei09a}.  It thus corresponds to the
infrared-to-bolometric luminosity ratio of the AGN.  A ULIRG powered
solely by an AGN, with no starburst contribution, would have a 100\% AGN fraction.
Solid symbols show data points derived from spectral fitting, and open symbols 
show data points estimated from counts (\S\ref{sec:specundet}).
To help guide the eye, we have
included two lines that show the expected values of the 2--10~keV luminosity ratio assuming a 1\% (bottom) and 10\% (top) AGN X-ray-to-bolometric correction. Of our sample, Mrk~231 and
IRAS~08572+3915 (assuming no intrinsic absorption) are particularly underluminous in the X-rays.  Note that Arp~220 is
excluded in this figure because we could not constrain its AGN properties.}
\label{fig:xfagn}
\end{figure}

\subsection{Comparison of hard X-ray and infrared perspectives}
\label{sec:irresults}

In Figure~\ref{fig:xfagn}, we compare the intrinsic 2--10~keV to bolometric
luminosity ratio, also known as the X-ray bolometric correction, to
the fraction of the bolometric luminosity attributed to the AGN from
infrared measurements.  This latter ratio, L$_{bol,AGN}$, is the average AGN fraction
calculated from six independent methods that include fine structure
line ratios, mid-infrared continuum ratios, and the equivalent
widths of the aromatic features \citep{vei09a}.
Included in the figure are also two lines that help guide the eye for
where pure AGN contributing 100\% of the luminosity in the infrared
with 1\% and 10\% X-ray bolometric correction would lie for a given
AGN fraction.

X-ray bolometric corrections of 2--15\% are typical for Seyferts and radio-quiet 
quasars \citep{elvis}, with much lower bolometric corrections of 
$\sim$0.3--0.7\%  seen for objects accreting at close to the Eddington rate
\citep[e.g.,][]{lusso10, lusso12, vf09}.   
While two of the eight ULIRGs lie within the range of bolometric corrections that
is typical for typical Seyferts, the majority appear surprisingly faint in the X-rays.  
Their X-ray bolometric corrections are more in line with those of 
objects accreting at close to the Eddington rate.

This disagreement between infrared and X-ray diagnostics is particularly 
large for  Mrk~231 and IRAS~08572+3915:  while 
 {\it Spitzer} diagnostics find that these two sources are heavily dominated by AGN, 
both are very underluminous in the X-rays. 
The ratio of intrinsic 2-10 keV luminosity to bolometric luminosity 
for Mrk~231 is only 0.03\%, and the ratio for IRAS~08572+3915 is even lower
 (assuming no intrinsic absorption.)
\citet{teng14} established that Mrk~231 is intrinsically X-ray weak,
likely related to the powerful wind  detected in this
broad-absorption line (BAL) quasar.   There is growing evidence
emerging that some AGN
with strong outflows, such as some BAL quasars, are intrinsically X-ray weak
\citep[e.g.,][]{luo13, luo14, teng14}, suggesting that intrinsic X-ray
weakness and  strong winds may be linked.  IRAS~08572+3915, known
to have a strong outflow on kiloparsec scales \citep{rv13}, may be
another example of an intrinsically X-ray weak AGN with powerful winds.
In fact, IRAS~08572+3915 is a more extreme case of intrinsic X-ray weakness than Mrk 231.  
\citet{Efstathiou14} 
identified IRAS~08572+3915 as the most infrared luminous galaxy, 
with its AGN contributing $\sim$90\% of its total power output.

Ultra-fast outflows have recently been discovered in 
two nearby ULIRGs, IRAS~F11119+3257  \citep{Tombesi2015} and Mrk~231 \citep{feruglio15},
that also host large-scale molecular  outflows.  
Although large scale outflows are quite different from the disk winds directly observed in BAL 
quasars, the presence of both kinds of outflows in Mrk~231 suggests that the 
two phenomena may be related.

\begin{figure}[h]
\centering
\includegraphics[width=3in]{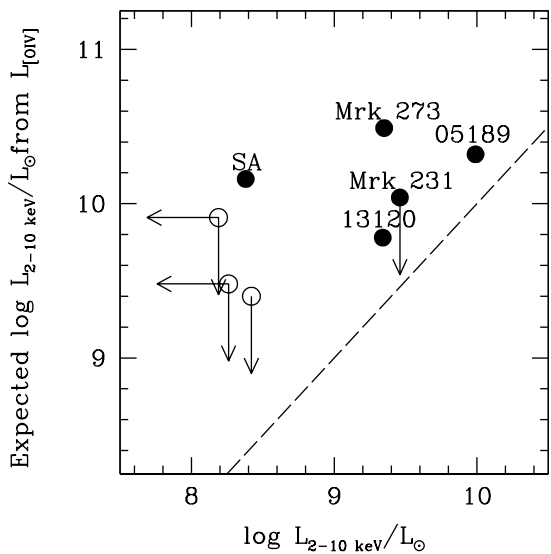}
\caption{A comparison of the intrinsic 2--10~keV luminosity
inferred from measured [O~IV] luminosity \citep{rigby09}, to the 
  intrinsic 2--10~keV luminosity as measured from broadband X-ray
  spectral modeling.
Solid symbols show data points derived from spectral fitting, and open symbols 
show data points estimated from counts (\S\ref{sec:specundet}).
  The dashed line is the one-to-one ratio to
   guide the eye.  The [O~IV] luminosity over-predicts the
  intrinsic 2--10~keV luminosity in all cases where both 
  quantities are detected.
}  
\label{fig:210comp}
\end{figure}

In Figure~\ref{fig:xfagn} we considered the fraction of bolometric luminosities attributable to the AGN.  
Now we consider diagnostics of the intrinsic AGN luminosities.  
One such method is the [O~IV]~26~\micron\ luminosity to 2--10~keV
luminosity relation \citep[e.g.,][]{melendez08, diamondstanic09,rigby09, weaver10}.  
[O IV]~26~\micron\ is far more robust to extinction than is the optical emission line [O III], 
and thus a more suitable diagnostic for highly obscured objects like ULIRGs, 
for which star formation also contributes significantly to their overall power.

We first investigate the possibility that the [O IV] line is
contaminated by star formation.   ULIRGs have significant circumnuclear star formation
\citep[see][for a detailed analysis on Mrk~231]{teng14}, and the hottest stars
are capable of creating some of the 55~eV photons required to 
triply ionize oxygen.   We investigate this possibility by comparing the [O IV] fluxes to 
[Ne V].  Even the hottest stars should not generate significant amounts of the 97~eV photons 
required to quadruply ionize neon within a galaxy; 
accordingly, [Ne V] is not detected in the mid-infrared
spectra of star-forming galaxies.  \citet{GouldingAlexander09} find a tight relation
between the [O IV]~26~\micron\ and [Ne V]~14.3~\micron\ emission in AGN, 
shown in their Figure 10, and argue that this is a method of determining whether
an AGN dominates the [O IV] luminosity.  
Four ULIRGs in our sample have both [O IV] and [Ne V] detected;
in the other objects neither line is detected.  These four objects lie 
close to the \citet{GouldingAlexander09} relation, at the upper right corner
with 41 $<$ log L[O IV] $<$ 42.2~erg~s$^{-1}$.
 If anything, these objects have somewhat higher [Ne V]/[O IV] ratios than the 
relation.  This argues that the [O IV] emission, like the [Ne V] emission, is powered
by the AGN, and that there is not significant [O IV] contamination from 
extremely hot stars.

We therefore proceed to 
compare the measured intrinsic 2--10~keV luminosities of our ULIRGs
with the expected intrinsic 2--10~keV luminosities predicted by the
measured [O~IV] luminosity and the \citet{rigby09} relation.
Figure~\ref{fig:210comp} shows that the [O~IV] relation, which was calibrated from
empirical measurements of Seyfert 1 galaxies, overpredicts the intrinsic 2--10~keV
luminosity for all sources in our sample.  We now discuss possible 
reasons for this discrepancy. 

First, it is worth noting that [O IV] was only detected in half the sample.  We attribute
this to the difficulty of detecting a weak line over a very strong continuum in the 
moderate signal-to-noise ratio {\it Spitzer} IRS spectra.  It is possible, though implausible, that 
the actual [O IV] fluxes for the non-detections are very low, enough to pull down 
these four sources to match the dotted line in Figure~\ref{fig:210comp}.

The second possibility is that the AGN
in some ULIRGs have different UV-to-X-ray ratios than Seyfert galaxies from
which the relation is derived.  The [O~IV] and [Ne V] emission arises in the
narrow line region,  where 55~eV and 97~eV (respectively) 
photons that originate in the accretion disk are able to triply ionize oxygen and 
quadruply ionize neon.
Thus, these high ionization fine structure lines are measures of the intrinsic UV luminosity.  
The X-ray luminosity is also tied to the accretion disk, as it is produced by
Compton upscattering of UV photons produced in the accretion disk by hot electrons in the corona. 
Thus, Figure~\ref{fig:210comp} suggests that the AGN in ULIRGs may have
different X-ray--producing efficiencies than do typical AGN in Seyfert galaxies, 
which we speculate might be due to differences in the structure or geometry of the corona.

To contextualize, the long effort to determine the AGN content of ULIRGs has involved  
two parallel paths:  extinction--robust hard X-ray photons, and extinction--robust mid-infrared 
signatures.   Both approaches assume that the AGN within a ULIRG has the properties of a
typical Seyfert AGN, and is merely highly obscured.  Our work reveals that these assumptions may break
down in certain regimes.
Objects like Mrk~231 and IRAS~08572+3915 have infrared signatures of AGN dominance, but are
severely under-luminous in the X-rays, and the cause is not high extinction.  
The rest of the sample appears X-ray under-luminous as well (see Figures~8 and 9.)
\citet{iman04} reached similar conclusions for IRAS~08572+3915 and three other
ULIRGs by comparing H$\alpha$ to {\it Chandra} and {\it XMM--Newton} spectra.
The ULIRGs in our sample have systematically low intrinsic 2-10~keV luminosities for
their measured [O IV]~26~\micron\ luminosities, compared to a relation calibrated with Seyfert~1 AGNs.
Both lines of evidence argue that the AGN within some, perhaps many ULIRGs 
may not be typical Seyfert 1 nuclei. 
We suggest that the violent conditions within some ULIRGs -- high accretion rates, strong winds -- may be
feeding back and affecting the X-ray  properties of their AGN.

\section{Summary}

We performed a detailed analysis of nine nearby ULIRGs observed by
{\it NuSTAR}.  The unprecedented sensitivity of {\it NuSTAR} at energies above 10~keV 
enables much improved constraints on the intrinsic X-ray properties of these galaxies.
Using 0.5--30~keV spectra from {\it NuSTAR}, {\it XMM--Newton}, and {\it Chandra}, 
we examined the hard X-ray properties of these presumed highly obscured sources.  We found that:

\begin{itemize}

\item The AGN within many ULIRGs are obscured  by column densities that are Compton-thin.
 Of the nine sources
observed by {\it NuSTAR}, six were detected well enough to allow
detailed spectral modeling of their broadband X-ray spectra.  Of
these six, only one (IRAS~13120--5453) has a spectrum consistent with
a Compton-thick AGN, though it is possible that Arp~220 is highly
Compton-thick ($N_H > 10^{25}$~cm$^{-2}$).  Only two of nine ULIRGs
have X-ray--luminous AGN with intrinsic 2--10~keV luminosities above
 $10^{43}$~erg~s$^{-1}$.

\item The ULIRGs in our sample have low ratios of intrinsic 2-10 keV luminosity to 
bolometric luminosity, below 1$\%$. This is much lower than the 
ratios of 2--15\% observed for Seyferts and radio-quiet
quasars \citep{elvis}, and closer to the ratios of
 $\sim$0.3--0.7\%  observed for objects with accretion rates close to the Eddington rate 
\citep{lusso10, lusso12, vf09}.   
The resulting low intrinsic X-ray luminosities have contributed 
to the lack of detection in past surveys, leading to previous conclusions that many ULIRGs are 
Compton-thick. 

\item IRAS~08572+3915 and Mrk 231 are likely intrinsically very weak in the X-rays.
The intrinsic X-ray weakness may be associated with
  powerful outflows, similar to broad-absorption line quasars. 

\item Established correlations between [O~IV] luminosity and intrinsic 2--10~keV
  luminosity, developed from samples of Seyfert galaxies, may not be appropriate for ULIRGs.

\end{itemize}

\acknowledgements We thank Lee Armus who provided useful
comments in the early planning phase of the {\it NuSTAR} ULIRG
program.  This work was supported under NASA Contract No. NNG08FD60C,
and  made use of data from the {\it NuSTAR} mission, a project led by
the California Institute of Technology, managed by the Jet Propulsion
Laboratory, and funded by the National Aeronautics and Space
Administration. We thank the {\it NuSTAR} Operations, Software and
Calibration teams for support with the execution and analysis of
these observations.  This research has made use of the {\it NuSTAR}
Data Analysis Software (NuSTARDAS) jointly developed by the ASI
Science Data Center (ASDC, Italy) and the California Institute of
Technology (USA).  The scientific results reported in this article are
based in part on observations made by the {\it Chandra X-ray
  Observatory} and data obtained from the {\it Chandra} Data Archive
published previously in cited articles.  This work, in part, made use
of observations obtained with {\it XMM-Newton}, an ESA science mission
with instruments and contributions directly funded by ESA Member
States and the USA (NASA).  We made use of the NASA/IPAC Extragalactic
Database (NED), which is operated by the Jet Propulsion Laboratory,
Caltech, under contract with NASA.  
S.H.T. was supported by a NASA Postdoctoral Program Fellowship.
Partial funding for this research was provided by a NASA XMM-Newton AO-12 
Grant award associated with proposal number 72261.
Support for the work of ET was provided by the Center of Excellence in 
Astrophysics and Associated Technologies (PFB 06), by the FONDECYT regular grant 
1120061 and by the CONICYT Anillo project ACT1101.

{\it Facilities:} \facility{{\it NuSTAR}, {\it Chandra}, {\it XMM-Newton}}.

\begin{turnpage}
\begin{deluxetable*}{lcccccl}
\tablecolumns{5}
\tabletypesize{\scriptsize}
\setlength{\tabcolsep}{0.01in}
\tablecaption{Best-fit Parameters for the ULIRGs in the Deep Survey}
\tablewidth{0pt}
\tablehead{\colhead{Model} & \colhead{IRAS 05189--2524} &
  \colhead{IRAS 13120--5453} & \colhead{Mrk~273}
  &\colhead{Arp~220}&\colhead{Superantennae}&\colhead{Comment on}\\
\colhead{Parameter} &&&&&&\colhead{Parameter}\\
\colhead{(1)}&\colhead{(2)}&\colhead{(3)}&\colhead{(4)}&\colhead{(5)}&\colhead{(6)}&\colhead{(7)}
}
\startdata
B/A&1.02$^{+0.03}_{-0.03}$&1.05 (f)&0.99$^{+0.06}_{-0.06}$&1.05
(f)&1.05 (f)&FPMB-A cross-normalization\\
{\it XMM}/A or {\it
  CXO}/A&0.86$^{+0.02}_{-0.02}$&1.20* (f) &0.88$^{+0.15}_{-0.14}$&0.90
(f)&0.85 (f)&
{\it XMM-Newton}- or {\it Chandra} (*)-   to FPMA cross-normalization\\
$k$T
[keV]&0.16$^{+0.01}_{-0.01}$&0.56$^{+0.05}_{-0.06}$&0.64$^{+0.15}_{-0.15}$&0.16$^{+0.02}_{-0.02}$,
0.69$^{+0.10}_{-0.10}$&0.51$^{+0.20}_{-0.17}$&MEKAL gas temperature from the starburst\\
Brems. $k$T [keV]&\nodata&\nodata&\nodata&9.13$^{+3.12}_{-2.04}$&\nodata&Bremsstrahlung temperature from the hot gas\\
Abs. 1 [$10^{22}$~cm$^{-2}$]&5.19$^{+0.20}_{-0.18}$&315.7$^{+232.5}_{-129.4}$&43.8$^{+9.5}_{-5.7}$&0.38$^{+0.04}_{-0.05}$&\nodata& neutral absorber 1\\
cf 1&0.98$^{+0.01}_{-0.01}$& 1 (f) &1 (f) & 1(f) &\nodata&covering factor 1\\
Abs. 2 [$10^{22}$~cm$^{-2}$]&9.32$^{+0.95}_{-0.68}$ &\nodata&\nodata& \nodata&\nodata&neutral absorber 2\\
cf 2&0.74$^{+0.01}_{-0.02}$&\nodata&\nodata&\nodata&\nodata&covering factor 2\\
Abs. HMXB [$10^{22}$~cm$^{-2}$]&$>
2.66$&0.58$^{+1.32}_{-0.58}$&\nodata&\nodata&\nodata&neutral absorber applied
to the HMXB power law component\\
$\Gamma_{\rm HMXB}$&1.1 (f)&1.1 (f)& 1.1 (f)&\nodata&1.1 (f) &HMXB cutoff power law index with cutoff energy at 10~keV\\
$\Gamma_{\rm AGN}$&2.51$^{+0.02}_{-0.02}$&1.8 (f)&1.43$^{+0.17}_{...}$&\nodata&1.54$^{+0.07}_{-0.07}$&AGN power law index (MYTorus lower limit fixed at 1.4)\\
Inc [$^\circ$]&\nodata&85 (f) &85 (f)&\nodata&\nodata&inclination angle\\
E$_{\rm
  line}$[keV]&6.43$^{+0.05}_{-0.05}$&1.86$^{+0.11}_{-0.05}$&\nodata&6.78$^{+0.07}_{-0.07}$&6.53$^{+0.16}_{-0.11}$&Line
1\\
EW$_{\rm line}$[keV]&0.074$^{+0.033}_{-0.035}$&0.104
(uc)&\nodata&0.899$^{+0.469}_{-0.398}$&0.296$^{+0.447}_{-0.107}$&EW of line 1\\
E$_{\rm line}$ 2 [keV]&6.80$^{+0.02}_{-0.04}$&3.40$^{+0.11}_{-0.06}$&\nodata&\nodata&
6.88$^{+0.37}_{-0.08}$&Line 2\\
EW$_{\rm line}$ 2
[keV]&0.117$^{+0.028}_{-0.040}$&0.460$^{+0.348}_{-0.279}$&\nodata&\nodata&
$0.330^{+0.538}_{-0.148}$& EW of line 2\\
E$_{\rm line}$ 3 [keV]&\nodata&6.86$^{+0.28}_{-0.12}$&\nodata&\nodata&
\nodata&Line 3\\
EW$_{\rm line}$ 3
[keV]&\nodata&0.848$^{+0.674}_{-0.562}$&\nodata&\nodata& \nodata& EW
of line 3\\
Const. (C-thin)&\nodata&\nodata&0.03$^{+0.02}_{-0.02}$&\nodata&\nodata&Compton-thin fraction\\
$f_{0.5-2}$ [$10^{-13}$ erg~s$^{-1}$~cm$^{-2}$]&$1.50^{+0.48}_{-0.50}$&0.50$^{+0.08}_{-0.24}$&$0.98^{+0.11}_{-0.14}$&0.98$^{+0.04}_{-0.04}$&0.75$^{+0.04}_{-0.05}$&observed 0.5--2 keV flux\\
$f_{2-10}$ [$10^{-12}$ erg~s$^{-1}$~cm$^{-2}$]&$3.87^{+0.11}_{-0.13}$&0.26$^{+0.30}_{-0.33}$&$0.76^{+0.03}_{-0.25}$&0.12$^{+0.01}_{-0.01}$&0.23$^{+0.01}_{-0.01}$&observed 2--10 keV flux\\
$f_{10-30}$ [$10^{-12}$ erg~s$^{-1}$~cm$^{-2}$]&$2.74^{+0.11}_{-0.16}$&0.89$^{+0.29}_{-0.78}$&$2.97^{+0.09}_{-1.40}$&0.05$^{+0.02}_{-0.02}$&0.25$^{+0.03}_{-0.03}$&observed 10--30 keV flux\\
L$_{\rm MEKAL}$ [erg~s$^{-1}$]&$1.40 \times
10^{41}$&$2.03\times10^{41}$&$1.47 \times 10^{41}$&$2.80 \times10^{41}$&$1.77\times10^{41}$&intrinsic MEKAL 0.5--30~keV luminosity\\
L$_{\rm Brems}$ [ergs~s$^{-1}$]&\nodata&\nodata&\nodata&$1.66\times10^{40}$&\nodata&intrinsic bremsstrahlung 0.5--30~keV luminosity\\
L$_{\rm HMXB}$ [erg~s$^{-1}$]&$4.70 \times 10^{41}$&$7.62\times10^{41}$&$7.09\times 10^{41}$&\nodata&$2.79\times10^{41}$&intrinsic HMXB 0.5--30~keV luminosity\\
L$_{0}$ [erg~s$^{-1}$]&$5.58\times 10^{43}$&$3.15\times10^{43}$&$2.40 \times 10^{43}$&\nodata&$4.57\times10^{42}$&intrinsic AGN 0.5--30 keV luminosity\\
L$_{0}$ (2--10~keV) [erg~s$^{-1}$]&$3.69\times 10^{43}$&$1.25\times10^{43}$&$8.55 \times 10^{42}$&\nodata&$1.70\times10^{42}$&intrinsic AGN 2--10 keV luminosity\\
L$_{\rm AGN}$/L$_{\rm bol, AGN}$ [\%]&0.81&0.67&0.28 &\nodata&0.08&2--10~keV X-ray-to-bolometric luminosity ratio for the AGN\\
\hline\\
$\chi^2$/d.o.f.&714.2/666&36.3/48&189.5/209&122.9/101&146.22/112&goodness of fit\\
\enddata
\tablecomments{(f) denotes a fixed parameter and (uc) detnotes an
  unconstrained parameter.  
Col.(1): Model parameter for each fit.  Col.(2):  Best-fit
model: Const.$\times$N$_{\rm H,
  Galactic}$ (MEKAL + Abs$_{\rm
  nuclear~HMXB}\times$cutoffPL$_{\rm
  nuclear~HMXB}$ + Abs$_{\rm 1}\times$Abs$_{\rm 2}\times$PL$_{\rm
  AGN}$ + Line(6.4~keV) + Line(6.7~keV))  Col.(3): Best-fit
model: Const.$\times$N$_{\rm H,
  Galactic}$ (MEKAL + Abs$_{\rm
  nuclear~HMXB}\times$cutoffPL$_{\rm
  nuclear~HMXB}$ + Line(6.7~keV) + MYTorus$\times$PL$_{\rm
  AGN}$)  Col.(4): Best-fit
model: Const.$\times$N$_{\rm H,
  Galactic}$ (MEKAL + cutoffPL$_{\rm
  nuclear~HMXB}$ + MYTorus$\times$PL$_{\rm
  AGN}$ + Const.$_{\rm C-thin}\times$PL$_{\rm AGN}$).  Col.(5): Best-fit
model: Const.$\times$N$_{\rm H,
  Galactic}\times$Abs$_{\rm 1}\times$ (MEKAL$_{1}$ + MEKAL$_{2}$ + zbremss + Line(6.7~keV)).  Col.(6): Best-fit
model: Const.$\times$N$_{\rm H,
  Galactic}$ (MEKAL + cutoffPL$_{\rm
  nuclear~HMXB}$ + Line(6.7~keV) + Line(6.9~keV) + PL$_{\rm
  AGN}$). Col.(7): Comments on model parameter. In the second row, the \textit{Chandra}-FPMA cross-normalization is 
marked with an asterisk; the others are for \textit{XMM-Newton}-FPMA.
}
\label{tab:fits}
\end{deluxetable*}
\end{turnpage}


\begin{thebibliography}{}
\bibitem[Aaronson \& Olszewski(1984)]{aaronson84} Aaronson, M. \&Olszewski, E.W. 1984, Nature, 309, 414
\bibitem[Adams \& Weedman(1972)]{aw72} Adams, T.F. \& Weedman, D.W. 1972, ApJ, 173, L109
\bibitem[Armus et al.(1989)]{armus89} Armus, L., Heckman, T.M., \& Miley, G.K. 1989, ApJ, 347, 727
\bibitem[Armus et al.(1990)]{armus90} Armus, L., Heckman, T.M., \& Miley, G.K. 1990, ApJ, 364, 471
\bibitem[Armus et al.(2007)]{armus} Armus, L., et al. 2007, ApJ, 656, 148
\bibitem[Braito et al.(2003)]{braito03} Braito, V., et al. 2003, A\&A,
  398, 107
\bibitem[Braito et al.(2004)]{braito} Braito, V., et al. 2004, A\&A, 420,79 
\bibitem[Braito et al.(2009)]{braito09} Braito, V., et al. 2009, A\&A,
  504, 53
\bibitem[Brightman \& Nandra(2011)]{bntorus} Brightman, M. \& Nandra,
  K. 2011, MNRAS, 413, 1206
\bibitem[Brightman et al.(2013)]{brightman13} Brightman, M. et
  al. 2013, MNRAS, 433, 2485
\bibitem[Bushouse et al.(2002)]{bushouse2002} Bushouse, H.~A., Borne, K.~D., Colina, L., et al.\ 2002, \apjs, 138, 1 
\bibitem[Caputi et al.(2007)]{caputi07} Caputi, K. et al. 2007, ApJ, 660, 97
\bibitem[Constantin et al.(2009)]{Constantin09} Constantin, A.,  Green, P., Aldcroft, T., et al.\ 2009, \apj, 705, 1336 
\bibitem[Dai et al.(2012)]{dai12} Dai, X., Shankar, F., \& Sivakoff, G.R., 2012, ApJ, 757, 180
\bibitem[de Grijp et al.(1987)]{degrijp} de Grijp, M.~H.~K., Lub, J., \& Miley, G.~K.\ 1987, \aaps, 70, 95 
\bibitem[Diamond-Stanic et al.(2009)]{diamondstanic09} Diamond-Stanic, 
A.~M., Rieke, G.~H., \& Rigby, J.~R.\ 2009, \apj, 698, 623 
\bibitem[Di Matteo et al.(2005)]{dimatteo} Di Matteo et al. 2005, Nature, 433, 604
\bibitem[DiPompeo et al.(2013)]{dipompeo13} DiPompeo, M.A., Brotherton, M.S., \& De Breuck, C. 2013, MNRAS, 428, 1565
\bibitem[Dickey \& Lockman(1990)]{nh} Dickey, J.M. \& Lockman, F.J. 1990, ARA\&A, 28, 215
\bibitem[Downes \& Solomon(1998)]{downes98} Downes, D. \& Solomon,
  P.M. 1998, ApJ, 507, 615
\bibitem[Downes \& Eckart(2007)]{downes07} Downes, D. \& Eckart,
  A. 2007, A\&A, 468, L57
\bibitem[Elbaz et al.(2010)]{Elbaz10} Elbaz, D. et al. 2010, A\&A, 518, L29
\bibitem[Elvis et al.(1994)]{elvis} Elvis, M., et al. 1994, ApJS, 95, 1
\bibitem[Efstathiou et al.(2014)]{Efstathiou14} Efstathiou, A., Pearson, C., Farrah, D., et al.\ 2014, \mnras, 437, L16 
\bibitem[Fabian et al.(1988)]{fabian88} Fabian, A.C., Done, C., \& Ghisellini, G., 1988, MNRAS, 232, 21
\bibitem[Farrah et al.(2003)]{farrah03} Farrah, D., Afonso, J., Efstathiou, A., Rowan-Robinson, M., Fox, M., \& Clements, D. 2003, MNRAS, 343, 585
\bibitem[Farrah et al.(2007)]{farrah07} Farrah, D., et al. 2007, ApJ,
  667, 149
\bibitem[Farrah et al.(2008)]{Farrah08} Farrah, D. et al. 2008, ApJ, 677, 957
\& Fiore, F. 2010, A\&A, 518, L155
\bibitem[Feruglio et al.(2015)]{feruglio15} Feruglio, C., Fiore, F., Carniani, S., et al.\ 2015, arXiv:1503.01481 
\bibitem[Franceschini et al.(2003)]{frances03} Franceschini, A., et al. 2003, MNRAS, 343, 1181
\bibitem[Gallagher et al.(2002)]{g02} Gallagher et al. 2002, ApJ, 569, 655
\bibitem[Gallagher et al.(2005)]{g05} Gallagher et al. 2005, ApJ, 633, 71
\bibitem[Gebhardt et al.(2000)]{gebhardt00} Gebhardt, K. et al. 2000, ApJ, 539, L13
\bibitem[Genzel et al.(1998)]{genzel98} Genzel, R., et al. 1998, ApJ,
  498, 579
\bibitem[Genzel \& Cesarsky(2000)]{genzelcesarsky00} Genzel, R. \& Cesarsky, C.J. 2000, ARA\&A, 38, 761
\bibitem[Genzel et al.(2001)]{genzel} Genzel, R., et al. 2001, ApJ, 563, 527
\bibitem[Gibson et al.(2009)]{gibson09} Gibson, R.R. et al. 2009, ApJ, 692, 758
\bibitem[Gibson \& Brandt(2012)]{Gibson12} Gibson, R.~R., \& Brandt, W.~N.\ 2012, \apj, 746, 54 
\bibitem[Giustini et al.(2008)]{giustini08} Giustini, M., Cappi, M.,
  \& vignali, C. 2008, A\&A, 491, 425
\bibitem[Gladders et al.(2013)]{Gladders13} Gladders, M.D. et al. 2013, ApJ, 764, 177
\bibitem[Gonz\'{a}lez-Alfonso et al.(2014)]{ga14} Gonz\'{a}lez-Alfonso, E. et al. 2014, A\&A, 561, 27
\bibitem[Goulding \& Alexander(2009)]{GouldingAlexander09} Goulding, A.~D., \& Alexander, D.~M.\ 2009, \mnras, 398, 1165 
\bibitem[Grupe et al.(2003)]{grupe03} Grupe, D., Mathur, S., \& Elvis,
  M. 2003, AJ, 126, 1159
\bibitem[Haan et al.(2011)]{haan11} Haan, S. et al. 2011, AJ, 141, 100
\bibitem[Harrison et al.(2013)]{nustar} Harrison, F.A. et al. 2013, ApJ, 770, 103
\bibitem[Hinshaw et al.(2009)]{cosmo} Hinshaw, G., Weiland, J.L., Hill, R.S., et al. 2009, ApJS, 180, 225
\bibitem[Hopkins et al.(2005)]{hopkins05} Hopkins, P.F. et al. 2005, ApJ, 630, 705
\bibitem[Hopkins et al.(2008)]{hopkins08} Hopkins, P.F. et al. 2008,
  ApJ, 175, 356
\bibitem[Hwang et al.(2010)]{Hwang10} Hwang, H.S. et al. 2010, MNRAS,
  409, 75
\bibitem[Imanishi \& Terashima(2004)]{iman04} Imanish, M. \&
  Terashima, Y. 2004, AJ, 127, 758
\bibitem[Iwasawa et al.(2005)]{iwasawa05} Iwasawa, K., Sanders, D.B.,
  Evans, A.S., Trentham, N., Miniutti, G., \& Spoon, H.W.W. 2005,
  MNRAS, 357, 565
\bibitem[Iwasawa et al.(2011)]{iwasawa11} Iwasawa, K. et al. 2011, A\&A, 529, 106
\bibitem[Iwasawa et al.(2012)]{iwasawa12} Iwasawa, K. et al. 2012, A\&A, 537, 86
\bibitem[Jia et al.(2012)]{jia12} Jia, J., Ptak, A., Heckman, T.M.,
  Braito, V., \& Reeves, J., 2012, ApJ, 759, 41
\bibitem[Kawakatu et al.(2007)]{kawakatu07} Kawakatu, N., Imanishi, M., \& Nagao, T. 2007, ApJ, 661, 660
\bibitem[Kennicutt(1998)]{ken98} Kennicutt, R.C. 1998, ApJ, 498, 541
\bibitem[Khachikian \& Weedman(1974)]{khachikian74} Khachikian, E.Y. \& Weedman, D.W. 1974, 192, 581
\bibitem[Kim \& Sanders(1998)]{ks98} Kim, D.-C. \& Sanders, D.B. 1998,
  ApJS, 119, 41
\bibitem[Kim et al.(1998)]{kim98} Kim, D.-C., Veilleux, Sylvain, \& Sanders, D.B. 1998, ApJ, 508, 627
\bibitem[Kim et al.(2002)]{kim02} Kim, D.-C., et al. 2002, ApJS, 143, 277
\bibitem[Koss et al.(2013)]{koss13} Koss, M. et al. 2013, ApJ, 765, L26
\bibitem[Krolik \& Kallman(1987)]{KrolikKallman87} Krolik, J.~H., \& Kallman, T.~R.\ 1987, \apjl, 320, L5 
\bibitem[Lebouteiller et al.(2011)]{cassis} Lebouteiller, V., Barry, D.~J., Spoon, H.~W.~W., et al.\ 2011, ApJS, 196, 8 
\bibitem[Le Floc'h et al.(2005)]{lefloch05} Le Floc'h, E., Papovich, C., Dole, H., et al.\ 2005, \apj, 632, 169 
\bibitem[Lehmer et al.(2010)]{lehmer} Lehmer, B.D. et al. 2010, ApJ, 724, 559
\bibitem[Levenson et al.(2002)]{Levenson02}  Levenson, N.~A., Krolik, J.~H., {\.Z}ycki, P.~T., et al.\ 2002, \apjl, 573, L8 
\bibitem[Luo et al.(2013)]{luo13} Luo, B. et al. 2013, ApJ, 772, 153
\bibitem[Luo et al.(2014)]{luo14} Luo, B. et al. 2014, ApJ, 794, 70
\bibitem[Lusso et al.(2010)]{lusso10} Lusso, E. et al. 2010, A\&A, 512, 34
\bibitem[Lusso et al.(2012)]{lusso12} Lusso, E. et al. 2012, MNRAS, 425, 623
\bibitem[Lynds(1967)]{lynds67} Lynds, C.R. 1967, ApJ, 147, 396
\bibitem[Madsen et al.(2015)]{Madsen15}  Madsen, K.~K., Harrison, F.~A., Markwardt, C.~B., et al.\ 2015, \apjs, 220, 8 
\bibitem[Markoff et al.(2005)]{markoff05} Markoff, S., Nowak, M.A., \& Wilms, J. 2005, ApJ, 635, 1203
\bibitem[Mart\'{i}nez-Sansigre et al.(2005)]{martinez05} Mart\'{i}nez-Sansigre, A., Rawlings, S., Lacy, M., Fadda, D., Marleau, F.R., Simpson, C., Willott, C.J., \& Jarvis, M.J., 2005, Nature, 436, 666
\bibitem[Melendez et al.(2008)]{melendez08} Melendez, M., Kraemer,
  S.B., Schmitt, H.R., et al. 2008, ApJ, 689, 95
\bibitem[Menendez-Delmestre et al.(2009)]{MenendezDelmestre09} Menendez-Delmestre, K. et al. 2009, ApJ, 699, 667
\bibitem[Mineo et al.(2012a)]{mineo12a} Mineo, S., Gilfanov, M., \& Sunyaev, R. 2012, MNRAS, 419, 2095
\bibitem[Mineo et al.(2012b)]{mineo12b} Mineo, S., Gilfanov, M., \&
  Sunyaev, R. 2012, MNRAS, 426, 1870
\bibitem[Miniutti et al.(2012)]{miniutti12} Miniutti, G., Brandt,
  N.W., Schneider, D.P. et al. 2012, MNRAS, 425, 1718
\bibitem[Murphy \& Yaqoob(2009)]{mytorus} Murphy, K.D. \& Yaqoob, T. 2009, MNRAS, 397, 1549
\bibitem[Murray et al.(1995)]{murray95} Murray, N., Chiang, J., Grossman, S.A., \& Voit, G.M. 1995, ApJ, 451, 498
\bibitem[Nagase(1989)]{hmxb} Nagase, F. 1989, PASJ, 41, 1
\bibitem[Naik et al.(2011)]{cenx3} Naik, S., Paul, B., \& Ali, Z. 2011, ApJ, 737, 79
\bibitem[Nandra \& Pounds(1994)]{nandra} Nandra, K. \& Pounds, K.A., 1994, MNRAS, 268, 405
\bibitem[Nandra et al.(2007)]{nandra07} Nandra, K. et al., 2007, MNRAS, 382, 194
\bibitem[Narayan \& Yi(1994)]{narayan94} Narayan, R. \& Yi, I., 1994, ApJ, 428, L13 
\bibitem[Narayan \& Yi(1995)]{narayan95} Narayan, R. \& Yi, I., 1995,
  ApJ, 452, 710
\bibitem[Neugebauer et al.(1984)]{neugebauer84} Neugebauer, G. et al. 1984, Science, 224, 14
\bibitem[Ogle et al.(1999)]{ogle99} Ogle, P.M., Cohen, M.H., Miller, J.S., et al. 1999, ApJS, 125, 1
\bibitem[Perez-Gonzalez et al.(2005)]{pg05} Perez-Gonzalez, P.G. et al. 2005, ApJ, 630, 82
\bibitem[Persic \& Raphaeli(2002)]{persic02} Persic, M. \& Raphaeli, Y. 2002, A\&A, 382, 843
\bibitem[Piconcelli et al.(2013)]{piconcelli} Piconcelli et al. 2013, MNRAS, 428, 1185 
\bibitem[Papadopoulos et al.(2012)]{papa12} Papadopoulos, P.P., van
  der Werf, P.P., Xilouris, E.M., Isaak, K.G., Gao, Y., \& M\"{u}hle,
  S., 2012, MNRAS, 426, 2601
\bibitem[Papovich et al.(2007)]{Papovich07} Papovich, C. et al. 2007, ApJ, 668, 45
\bibitem[Ptak et al.(2003)]{ptak03} Ptak, A. et al. 2003, ApJ, 592, 782
\bibitem[Reeves \& Turner(2000)]{reeves} Reeves, J.N. \& Turner, M.I.J., 2000, MNRAS, 316, 234
\bibitem[Reynolds et al.(2013)]{reynolds13} Reynolds, C., Punsly, B.,
  O'Dea, C.P., \& Hurley-Walker, N., 2013, ApJ, 776, L21
\bibitem[Rigby et al.(2008)]{Rigby08} Rigby, J.R. et al. 2008, ApJ, 675, 262
\bibitem[Rigby et al.(2009)]{rigby09} Rigby, J.R., Diamond-Stanic,
  A.M., \& Aniano, G., 2009, ApJ, 700, 1878
\bibitem[Ross \& Fabian(2005)]{reflion} Ross, R.R. \& Fabian,
A.C. 2005, MNRAS, 358, 211
\bibitem[Rowan-Robinson et al.(2004)]{Rowan-Robinson04} Rowan-Robinson, M. et al. 2004, MNRAS, 351, 1290
\bibitem[Rowan-Robinson et al.(2005)]{Rowan-Robinson05}
  Rowan-Robinson, M. et al. 2005, AJ, 129, 1183
\bibitem[Rujopakarn et al.(2011)]{Rujopakarn11} Rujopakarn, W., Rieke, G.H., Eisenstein, D.J., \& Juneau, S. 2011, ApJ, 276, 93
\bibitem[Rupke et al.(2005)]{rupke05} Rupke, D.S., Veilleux, S., \& Sanders, D.B. 2005, ApJ, 632, 751
\bibitem[Rupke \& Veilleux(2011)]{rv11} Rupke, D.N.S. \& Veilleux, S., 2011, ApJ, 729, L27
\bibitem[Rupke \& Veilleux(2013)]{rv13} Rupke, D.N.S. \& Veilleux, S., 2013, ApJ, 768, 75
\bibitem[Sajina et al.(2006)]{Sajina06} Sajina, A. et al. 2006, MNRAS, 369, 939
\bibitem[Sanders et al.(1988)]{sanders88} Sanders, D.B. et al. 1988, ApJ, 328, L35
\bibitem[Sanders et al.(2003)]{rbgs} Sanders, D.B., Mazzarella, J. M.,
  Kim, D.-C., Surace, J.A., \& Soifer, B.T., 2003, AJ, 126, 1607
\bibitem[Scoville et al.(2000)]{scoville00} Scoville, N.Z., et al. 2000, AJ, 119, 991
\bibitem[Scoville et al.(2014)]{scoville14} Scoville, N., Sheth, K., Walter, F., et al.\ 2014, arXiv:1412.5183 
\bibitem[Shemmer et al.(2005)]{shemmer05} Shemmer, O., Brandt, W.N., Gallagher, S.C., et al. 2005, AJ, 130, 2522
\bibitem[Shemmer et al.(2008)]{shemmer08} Shemmer, O., Brandt, W.N., Netzer, H., Maiolino, R., \& Kaspi, S., 2008, ApJ, 682, 81
\bibitem[Smith et al.(2007)]{pahfit} Smith, J.D.T., Draine B.T., et al., 2007, ApJ, 656, 770
\bibitem[Springel et al.(2005)]{springel} Springel et al. 2005, Nature, 435, 629
\bibitem[Strickland \& Heckman(2007)]{strickland07} Strickland, D.K. \& Heckmand, T.M. 2007, ApJ, 658, 258
\bibitem[Surace et al.(1998)]{surace98} Surace, J.A., Sanders, D.B.,
  Wacca, W.D., Veilleux, S., \& Mazzarella, J.M. 1998, ApJ, 492, 116
\bibitem[Surace et al.(2000)]{surace00} Surace, J.A., Sanders, D.B., \& Evans, A.S., 2000, ApJ, 529, 170
\bibitem[Symeonidis et al.(2009)]{Symeonidis09} Symeonidis, M. et
  al. 2009, MNRAS, 397, 1728
\bibitem[Taniguchi et al.(1999)]{taniguchi99} Taniguchi, Y., Yoshino, A., Ohyama, Y., \& Nishiura, S. 1999, ApJ, 514, 660
\bibitem[Teng et al.(2005)]{teng05} Teng, S.H. et al. 2005, ApJ, 633, 664
\bibitem[Teng et al.(2008)]{teng08} Teng, S.H. et al. 2008, ApJ, 674, 133
\bibitem[Teng et al.(2009)]{teng09} Teng, S.H. et al. 2009, ApJ, 691, 261
\bibitem[Teng \& Veilleux(2010)]{teng10} Teng, S.H. \& Veilleux, S. 2010, ApJ, 725, 1848
  Baker, A.J. 2013, ApJ, 765, 95
\bibitem[Teng et al.(2014)]{teng14} Teng, S.H. et al. 2014, ApJ, 785, 19
\bibitem[Tombesi et al.(2015)]{Tombesi2015} Tombesi, F., Melendez, M., Veilleux, S., et al.\ 2015, arXiv:1501.07664 
\bibitem[Vasudevan \& Fabian(2007)]{vf07} Vasudevan, R.V. \& Fabian, A.C. 2007, MNRAS, 381, 1235
\bibitem[Vasudevan \& Fabian(2009)]{vf09} Vasudevan, R.V. \& Fabian, A.C. 2009, MNRAS, 392, 1124
\bibitem[Verner et al.(1996)]{verner} Verner, D.A., Ferland, G.J., Korista, K.T., \& Yakovlev, D.G. 1996, ApJ, 456, 487
\bibitem[Veilleux et al.(1995)]{vei95} Veilleux, S., Kim, D.-C.,
  Sanders, D.B., Mazzarella, J.M., \& Soifer, B.T., 1995, ApJS, 98,
  171
\bibitem[Veilleux et al.(1999a)]{vei99a} Veilleux, S., Kim, D.-C., \&
Sanders, D.B. 1999a, \apj, 522, 113
\bibitem[Veilleux et al.(1999b)]{vei99b} Veilleux, S., Sanders, D.B.,
\& Kim, D.-C. 1999b, \apj, 522, 139
\bibitem[Veilleux et al.(2009a)]{vei09a} Veilleux, S. et al. 2009a, ApJS, 182, 628 
\bibitem[Veilleux et al.(2009b)]{vei09b} Veilleux S. et al. 2009b, ApJ, 701, 587
\bibitem[Veilleux et al.(2013a)]{vei13a} Veilleux, S. et al. 2013a, ApJ, 764, 15
\bibitem[V{\'e}ron-Cetty \& V{\'e}ron(2010)]{vv2010} V{\'e}ron-Cetty, M.-P., \& V{\'e}ron, P.\ 2010, \aap, 518, AA10 
\bibitem[Weaver et al.(2010)]{weaver10} Weaver, K.A., Melendez, M., Mushotzky, R.F. et al. 2010, ApJ, 716, 1151
\bibitem[Wik et al.(2014)]{wik14} Wik, D., et al. 2014, ApJ, 792, 48
\bibitem[Wilms et al.(2000)]{wilms} Wilms, J., Allen, A., \& McCray, R. 2000, ApJ, 542, 914
  al. 2011, ApJ, 736, 28
\bibitem[Xia et al.(2002)]{mrk273x} Xia, X.Y., Xue, S.J., Mao, S.,
  Boller, Th., Deng, Z.G., \& Wu, H., 2002, ApJ, 564, 196
\bibitem[Yang et al.(2014)]{yang14} Yang, Q.-X. et al. 2014, MNRAS, 447(2), 1692–-1704.
\bibitem[Yaqoob(2012)]{yaqoob12} Yaqoob, T. 2012, MNRAS, 423, 3360
\end{thebibliography}
\end{document}